\pdfoutput=1
\documentclass[12pt,preprint]{aastex}
\usepackage{graphicx}
\usepackage{natbib}
\usepackage{amsmath}
\begin{document}
\tighten

\title{When Shock Waves Collide}

\author{
	P. Hartigan \altaffilmark{1},
        J. Foster \altaffilmark{2},
        A. Frank \altaffilmark{3},
        E. Hansen\altaffilmark{3},
	K. Yirak \altaffilmark{4}, 
	A. S. Liao \altaffilmark{1}
        P. Graham \altaffilmark{2}
        B. Wilde \altaffilmark{5},
        B. Blue \altaffilmark{6},
        D. Martinez \altaffilmark{7}
        P. Rosen \altaffilmark{2}
        D. Farley \altaffilmark{8}
        R. Paguio \altaffilmark{6}
	}

\vspace{1.0cm}

\altaffiltext{1}{Rice University, Department of Physics and Astronomy, 6100 S. Main, Houston, TX 77005-1892} 
\altaffiltext{2}{AWE, Aldermaston, Reading Berkshire, RG7 4PR, UK}
\altaffiltext{3}{University of Rochester, Department of Physics and Astronomy, Rochester, NY 14627-0171}
\altaffiltext{4}{Formerly at Los Alamos National Laboratory, Los Alamos, NM 87545}
\altaffiltext{5}{Los Alamos National Laboratory, Los Alamos, NM 87545}
\altaffiltext{6}{General Atomics, 3550 General Atomics Court, San Diego, CA 92121-1122}
\altaffiltext{7}{Lawrence Livermore National Laboratory, 7000 East Avenue, Livermore CA 94550}
\altaffiltext{8}{Sandia National Laboratories, 7011 East Ave. Livermore, CA 94550}

\begin{abstract}
Supersonic outflows from objects as varied as stellar jets, massive stars and novae often
exhibit multiple shock waves that overlap one another. When the intersection angle between two shock
waves exceeds a critical value, the system reconfigures its geometry to create a normal shock 
known as a Mach stem where the shocks meet. Mach stems are important for
interpreting emission-line images of shocked gas because a normal shock produces higher
postshock temperatures and therefore a higher-excitation spectrum than an oblique one does.
In this paper we summarize the results of a series of numerical simulations and 
laboratory experiments designed to quantify how Mach stems behave in supersonic plasmas
that are the norm in astrophysical flows.  The experiments test analytical
predictions for critical angles where Mach stems should form, and quantify how
Mach stems grow and decay as intersection angles between the incident shock and
a surface change.  While small Mach stems are destroyed by surface
irregularities and subcritical angles, larger ones persist in these situations,
and can regrow if the intersection angle changes to become more favorable.  The experimental and numerical
results show that although Mach stems occur only over a limited range of
intersection angles and size scales, within these ranges they are relatively robust, and hence
are a viable explanation for variable bright knots observed in HST images at the intersections
of some bow shocks in stellar jets.  


\end{abstract}

\section{Introduction}
\label{sec:intro}

Supersonic flows in astrophysics often contain multiple shock fronts that
form as a result of unsteady outflows. 
Examples include supernovae remnants \citep{Chevalier75,Anderson94},
shells of novae \citep{shara97}, and Herbig-Haro (HH) bow shocks
\citep{hartigan01}. High-resolution images of these objects
sometimes show bright knots where the shock fronts intersect, and the
motion of these knots differs from the overall motion of their parent shocks
\citep{hartigan11}.  It is important to understand the different types of geometries that
are possible at intersection points of shocks because the postshock temperature,
and therefore the line excitation, depends upon the orientation of the shock
relative to motion of the incident gas. In addition, if a bright knot in
a jet merely traces the intersection point, then its observed proper motion 
will follow the location of that point and will not represent the bulk motion
of the shock wave. Even bow shocks that appear smooth in ground-based images
can be affected by this phenomenon. For example, the stellar jet bow shocks HH~1 and HH~47 both
exhibit variable filamentary structure in the highest-resolution HST images that most likely
arises from irregularities in the shock surfaces \citep{hartigan11}.
Hence, to interpret both the emission line structure and
the observed kinematics of astrophysical shock waves accurately we
must understand the physics of shock intersections. 

\begin{figure} %fig1
\centering
\includegraphics[angle=0,scale=1.00]{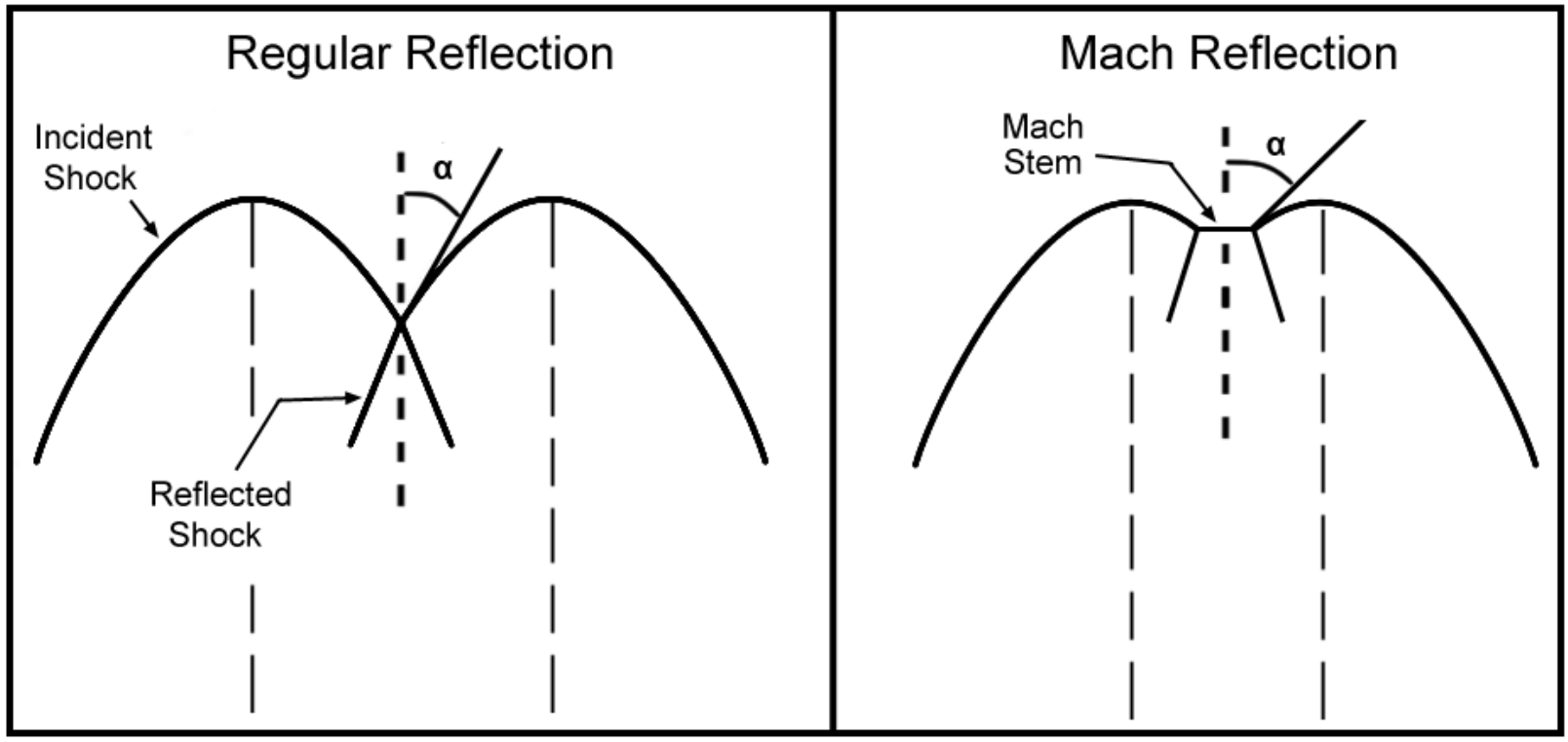}
\caption{Geometry of intersecting bow shocks.  Left: When the angle of intersection $\alpha$
is smaller than a critical value that depends on the specific heat ratio $\gamma$ of the
gas, regular reflection occurs and a single reflected shock follows in the wake
of the intersection point. Right: As the apices of the bow shocks approach one-another,
the system transitions to Mach reflection, and the intersection point
becomes a normal shock known as a Mach stem.
}
\label{fig:MS-cartoon}
\end{figure}

The theory of intersecting shock fronts is complex and has been studied for
over a century.  Modern reviews such as \cite{SWRF2nd} summarize a variety of shock
geometries that arise when shock waves collide.  
Early theoretical work by \citet{vn43} laid the foundation for the field
by identifying two general classes of solutions to the hydrodynamic equations at shock intersections
that he labeled as regular reflection (RR) and Mach reflection (MR; Fig.~\ref{fig:MS-cartoon}).
Regular reflection occurs in the case of intersecting bow shocks when the apices of two bows
are far apart and the shock waves intersect at an acute angle in the
wings of the bows. As shown in Fig.~\ref{fig:MS-cartoon}, by
symmetry, the gas that lies along the axis between the two bows will have no
net lateral deflection. Hence, a pair of reflected shocks must form behind the
intersection point between the bows that redirects the flow along the
axis of symmetry.

\begin{figure} %fig2
\centering
\includegraphics[angle=0,scale=1.00]{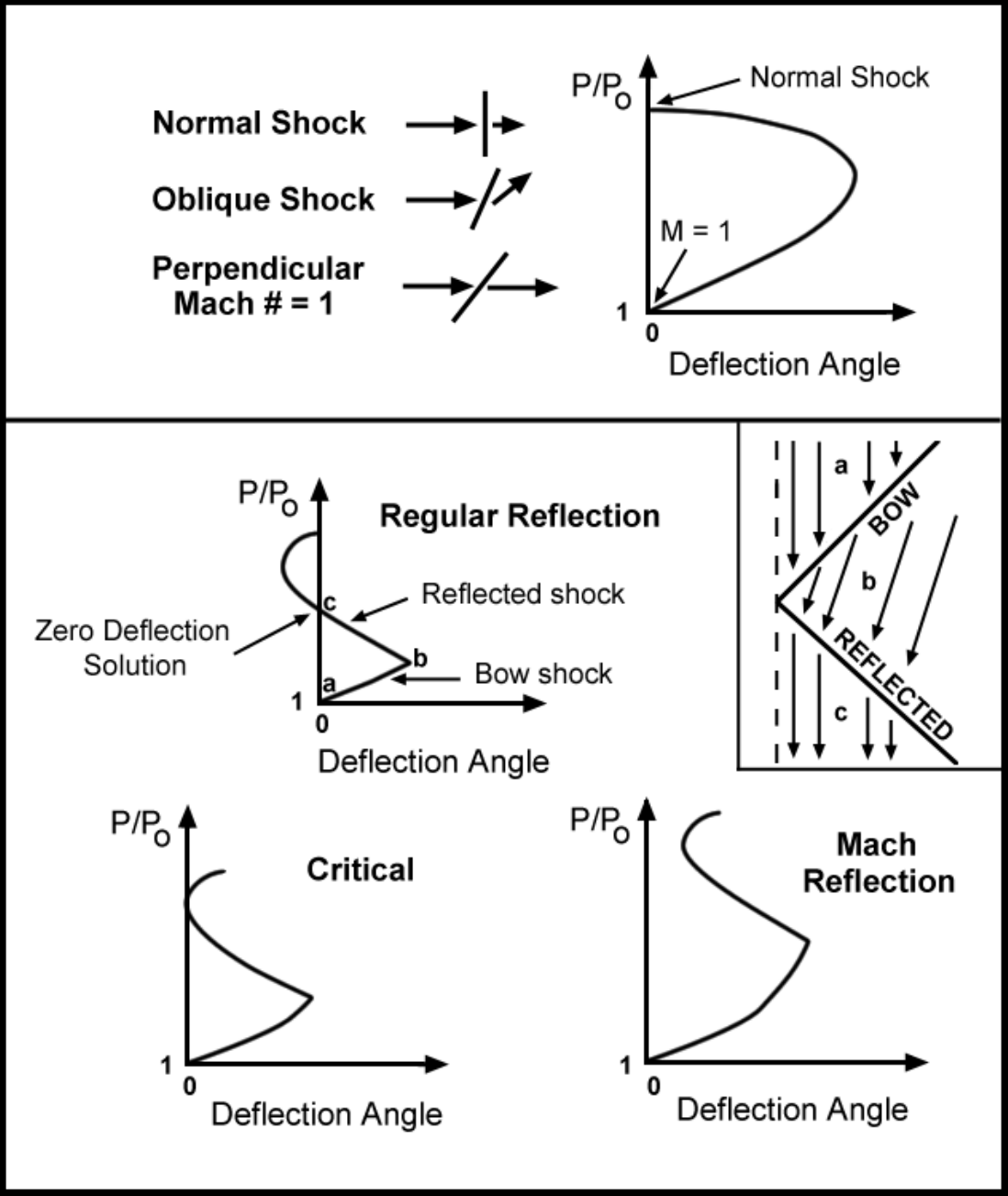}
\caption{Shock polars. Top: As the incident angle of the preshock gas
to a planar shock increases, the deflection angle first increases and then decreases.
The ratio of the postshock to preshock pressures is denoted as P/P$_\circ$.  Bottom:
Two oblique shock waves that deflect gas in opposite directions may either generate
two, one, or zero solutions that cross the y-axis in a polar diagram. These cases
correspond to regular reflection, critical, and Mach reflection, respectively. The inset
depicts regular reflection.
}
\label{fig:polar}
\end{figure}

As shown in the top panel of Fig.~\ref{fig:polar}, the amount that a planar shock deflects
the incident gas depends upon the angle of the shock to the flow.
Normal shocks produce zero deflection, and shocks that are inclined at the Mach angle,
defined so the perpendicular component of the velocity relative
to the front has a Mach number of 1, also do
not deflect the incident gas. At intermediate angles the shock deflects the gas away from the
shock normal, and the deflection goes through a maximum.
Plots of the deflection angle as a function of the effective Mach number
squared (or the ratio of the postshock pressure to the preshock pressure) are known as
shock polars \citep{ks56}.  As depicted in Fig.~\ref{fig:polar}, one can determine
the net amount of deflection from two planar shocks (in this case
the bow shock and its reflection shock) by attaching two polar curves together at the point 
where the two shocks intersect.  As one moves up along the reflected shock curve
in Fig.~\ref{fig:polar} it is the equivalent of changing the orientation angle
of the reflected shock.

When the two bow shocks are widely-separated, their intersection point lies far into
the wings of the bows. Material entering the bow shocks near this point encounters
a weak shock with a small deflection angle. The bottom panel of Fig~\ref{fig:polar}
shows that when the bow shock is weak (P/P$_\circ$ $\sim$ 1), the
reflected shock polar crosses the x-axis at two locations, implying there are
two solutions with zero net deflection \citep[in practice, systems choose the smaller
value that corresponds to the weaker shock;][]{SWRF2nd}. However, if the bow shocks are close together, the 
intersection angle between the shocks becomes more oblique, and a critical point
is reached where there is no solution for zero deflection. At more oblique angles
than the critical point, the system reconfigures to Mach reflection
(Fig.~\ref{fig:MS-cartoon}), consisting of the bow shock, reflected shock, and
a Mach stem that intersect at a triple point. Eventually, as the apices of the bow
shocks approach one-another the lateral motion behind the bow shocks becomes subsonic and
the triple point goes away, leaving the system with single bow shock. 
Theoretically, \cite{cf48} give an equation for the critical angle in the limit
of high Mach numbers as 

\begin{equation}
\label{eqn:cfeqn}
\alpha_{C} = sin^{-1}\left(\frac{1}{\gamma}\right)
\end{equation}  

\noindent
where $\gamma$ is the adiabatic index of the gas. A similar, but more detailed analytical expression
was derived by \cite{derosa92}. 

When the interaction angle rises above the critical angle, the RR should transition
to MR, and the system should return to RR if the critical angle decreases.
Theoretically, both MR and RR may occupy the same parameter space
\citep{lbd96}, and there is experimental evidence that the mode present in a given situation
depends upon the previous history of the gas, a phenomenon referred to as hysteresis.  
As described by \cite{Bendor2002} and \cite{chpoun95}, the critical angle for RR $\rightarrow$ MR 
differs slightly from the critical angle for MR $\rightarrow$ RR. In addition,
under some circumstances the region near the triple point can become more complex,
and even split into dual triple points known as double Mach-reflection \citep[hereafter DMR;][]{white51,hg86}.
These DMR structures involve rather small changes to the overall shock geometry compared with
the normal Mach stem case and currently have no obvious astrophysical analogs.

Experimentally, most of the effort to date regarding the MR $\leftrightarrow$ RR transition
has been directed to studying relatively low Mach number flows.
Exploring the physics of Mach stems at high Mach numbers has recently become
possible on experimental platforms that use intense lasers to drive strong shocks. 
One such platform, developed by \cite{foster10} on the
University of Rochester's Omega laser, drove pairs of strong shocks into a foam target. 
Temperatures achieved at the shock front were high
enough to ionize the medium, allowing a closer approximation to the
astrophysical plasma regime than shock tube studies have done in the past. 
In their experiment, \cite{foster10} formed 
Mach stems as two directly-opposed bow shocks collided, and  
measured the critical angle for MR to be 48$^{\circ}\pm$15$^{\circ}$. This
critical angle lies roughly halfway between those found from numerical simulations
of the experiment with $\gamma$ = 5/3 and $\gamma$ = 4/3. 

In this paper we extend the laser experimental work of \cite{foster10}
to a platform where we control the intersection angle between a strong shock
and a surface by constructing targets with shapes designed to (a) keep the intersection
angle constant, (b) decrease the intersection angle suddenly to below the critical
value, and then gradually increase it above that value and (c)
compare shock propagation over smooth and rough surfaces.
Initial results of the work were published by \cite{yirak13}. 
Here we compile all the data from two
years of experiments at the Omega laser facility, and supplement the experimental
data with a series of numerical simulations of intersecting shocks. 
The results provide a foundation we can use to assess the viability and importance
of Mach stems for interpreting images of shock fronts within astrophysical objects.

We describe the experimental setup and targets in \S2.  Results of the experiments,
described in section \S3, include new measurements of critical angles,
growth rates, and survivability of Mach stems
within inhomogeneous environments.  In \S4 we use numerical simulations
to clarify the physics of Mach stem evolution, and consider the broader
astrophysical implications of the research. \S5 presents a summary.

\section{Experimental Design} 
\label{sec:expr}

\begin{figure} %fig3
\centering
\includegraphics[angle=0,scale=1.00]{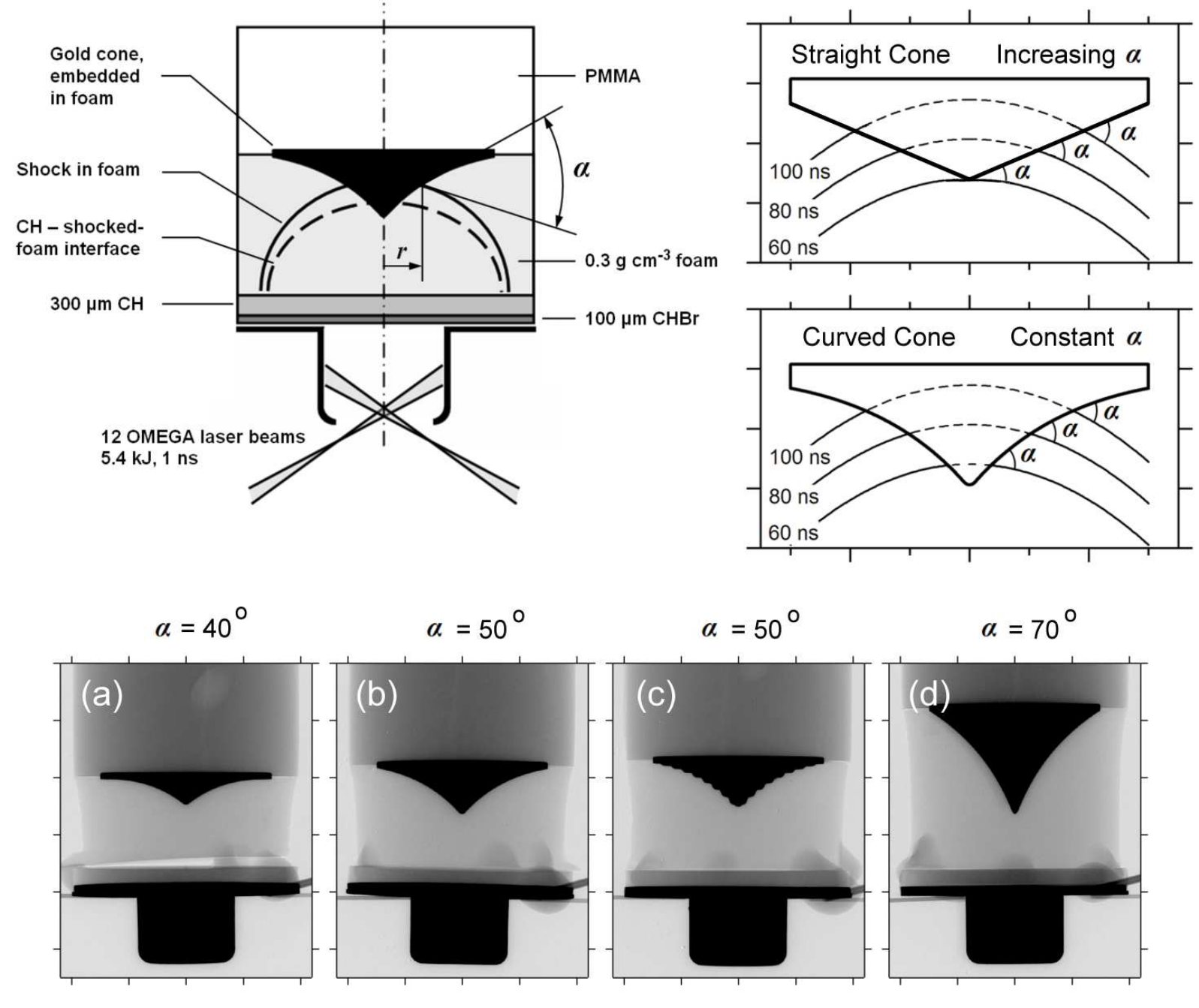}
\caption{Top left: Laser target design indicating the materials used and 
scale sizes. The base diameter of the cone is 3~mm.
Top right: Depiction of a curved bow shock moving along a straight cone (top)
and a curved cone (bottom).  The position of the bow shock is shown at
sequential times. The intersection angle $\alpha$ increases steadily with time
for the straight cone, and is constant for the curved cone. Tick marks are spaced
0.5~mm apart.
Bottom: Radiographs of curved cones that produce a constant value of $\alpha$ of
(a) 40$^\circ$, (b) and (c) 50$^\circ$, and (d) 70$^\circ$. The serrated
cone in panel (c) has a sinusoidal modulation. Tick marks are 1~mm apart.}
\label{fig:targets}
\end{figure}

The design of the laser experiments is summarized in Fig.~\ref{fig:targets}. 
We used the Omega laser \citep{omega} at the University of Rochester's Laboratory
for Laser Energetics to drive a strong curved shock through a cylinder of
low-density foam within which we embedded a cone-shaped obstacle.
In the frame of reference of the intersection point between the curved shock and the
cone, the surface of the cone provides a reflecting boundary condition identical to that
of two intersecting bow shocks.  Hence, our experimental design allows us to explore
this intersection point under controlled conditions.  The experiments used
indirect laser drive to launch the shock wave. Twelve laser beams, each with energy 450~J in a
1~ns pulse, impinge upon the inside surface of a hollow gold hohlraum laser target. The hohlraum had a
diameter of 1.6~mm, length 1.2~mm, and laser entrance hole of diameter 1.2~mm. The hohlraum converts
the laser energy to thermal X-rays, which subsequently ablate the surface of a composite ablator-pusher that acts
as a piston to drive a shock wave into the foam.

We used two types of ablator-pushers. In the
first experiments in this series, the ablator was a 100  $\mu$m thick layer
of CH doped with 2\% by-atom Br with density 1220 mg/cc. This was in contact with a 300
$\mu$m layer of CH (polystyrene) with density 1060 mg/cc that served as a pusher to propel a shock
wave into the foam. In later experiments, we changed the design of the ablator-pusher,
to reduce radiation preheat of the embedded cone. This design incorporated a 100 $\mu$m thickness CH (polystyrene)
ablator and 300 $\mu$m thickness Br-doped CH pusher (6\% by-atom Br, 1.5 g/cc density). In all cases, the
foam was resorcinol-formaldehyde of density 300 mg/cc. The foam and embedded cone (made from gold, because of its
high density) were supported from the end face of a polymethyl methacrylate (PMMA) cylinder of full density.

After a predetermined delay time that ranged from 50~ns to 150~ns to allow the shock wave to propagate
to the desired position, the shock wave was imaged by point-projection X-ray radiography along two mutually orthogonal
lines of sight, perpendicular to the axis of symmetry of the experimental assembly. The near-point X-ray sources for
backlit imaging were provided by using two further laser-illuminated targets, and the images were recorded on a
pair of time-gated micro-channel-plate X-ray framing cameras. The output optical images from these cameras were recorded
on either photographic film or with a CCD. The laser pulse duration for the x-ray backlighting sources was 600~ps.
Contrast in the radiographs is caused by transparency differences through the target, and can be adjusted by
altering the composition of the materials. For example, brominated CH absorbs X-rays much more strongly than pure
CH foam or plastic do, and will appear correspondingly darker in the radiographs. We used nickel disk laser
targets of 400 $\mu$m diameter and 5 $\mu$m thickness as backlighters. These were mounted on 5~mm square,
50 $\mu$m thickness pieces of tantalum sheet. A pinhole of diameter 10 or 20 $\mu$m was machined into
the centers of these tantalum supports to aperture x-ray emission from the nickel laser targets, thus providing a
near-point X-ray source. A time delay of 7~ns between the two X-ray backlighting laser sources 
makes it possible to produce two snapshot images, along orthogonal lines of sight and at different times, for
each target. The shock velocity through the foam at the position of the cone was typically 20 km$\,$s$^{-1}$.
Radiation preheated the foam to several hundred degrees C before the arrival of the shock, so the preshock
sound speed is $\sim$ 1 km$\,$s$^{-1}$ and the Mach number $\sim$ 20.

\begin{figure}  %fig4
\centering
\includegraphics[angle=0,scale=1.00]{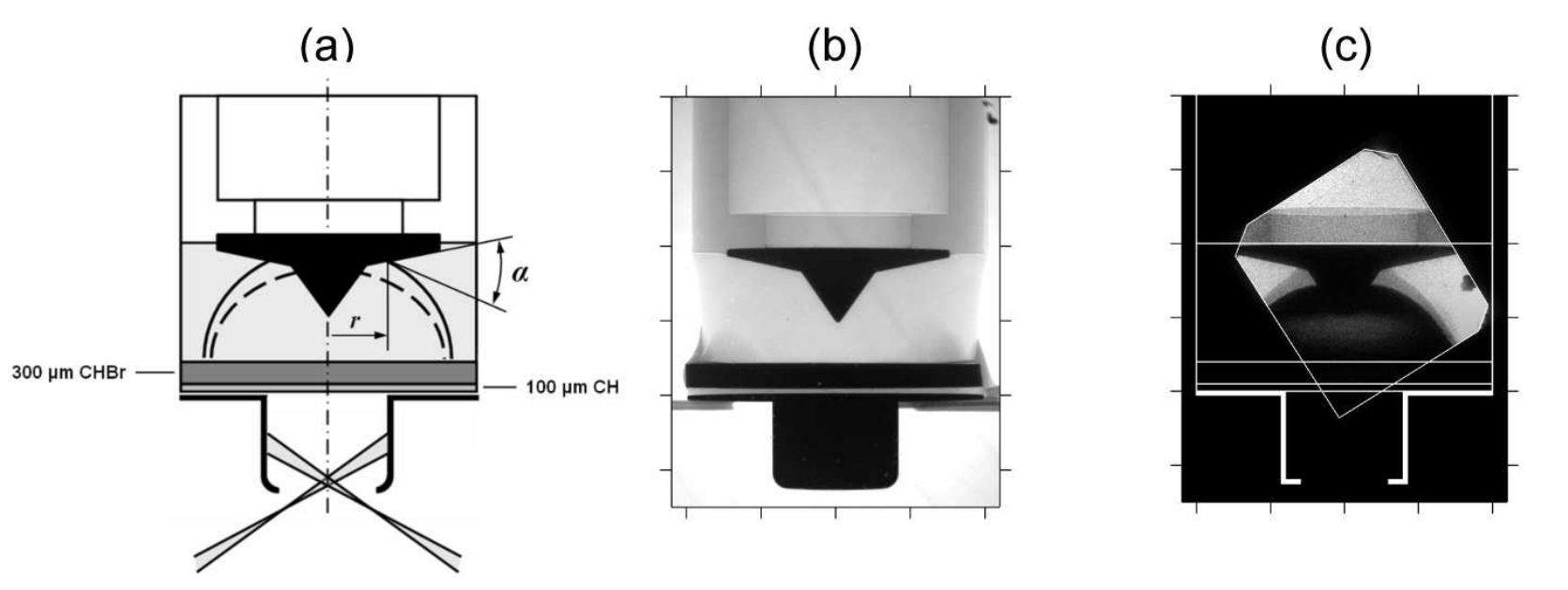}
\caption{
Dual-angle targets. (a) Target design, same as that in Fig.~\ref{fig:targets},
but with dual-cone half angles of 35$^\circ$ and 78$^\circ$. Cones with half
angles of 35$^\circ$ and 84$^\circ$ were also used. (b) Target radiograph.
(c) Radiograph of the shock as it interacts with the conical surface. The 
base diameter of the cone is 3~mm and tick marks in (b) and (c) are 1~mm
apart. The rectangular area in (c) indicates the diagnostic field of view.
}
\label{fig:dual}
\end{figure}

As depicted in Fig.~\ref{fig:targets}, the angle of intersection ($\alpha$) between a curved bow
shock and a straight-sided cone that has triangular cross section increases
steadily with time, while a curved-sided cone of appropriate form provides a constant
angle of intersection. 
To investigate Mach stem development at a fixed intersection angle we used
the concave cone designs shown at the bottom of Fig.~\ref{fig:targets} for
$\alpha$ = 40$^\circ$, 50$^\circ$, 60$^\circ$ and 70$^\circ$.  A dual-angled
cone like the one shown in Fig.~\ref{fig:dual} makes it possible to reduce the angle
of intersection abruptly to below the critical value, a transition that occurs at a radial distance of 400
$\mu$m or 500 $\mu$m from the cone axis for the targets that we used. Because the
dual-angle cones have flat conical cross sections beyond the transition point,
$\alpha$ once again increases to greater than the critical angle
at later times. Hence, these targets are ideal for investigating the decay and growth rates of Mach stems.
A final target design employed a constant incident-angle (50$^\circ$) cone
that was ‘terraced’ (panel c in Fig.~\ref{fig:targets}) in order to study
the degree to which Mach stems are disrupted by surface irregularities. Terracing of the cone was obtained by
machining a sinusoidal modulation onto its surface. Results from
these experiments are discussed in \S\ref{sec:bumpy}.

\begin{figure}  %fig5
\centering
\includegraphics[angle=0,scale=1.00]{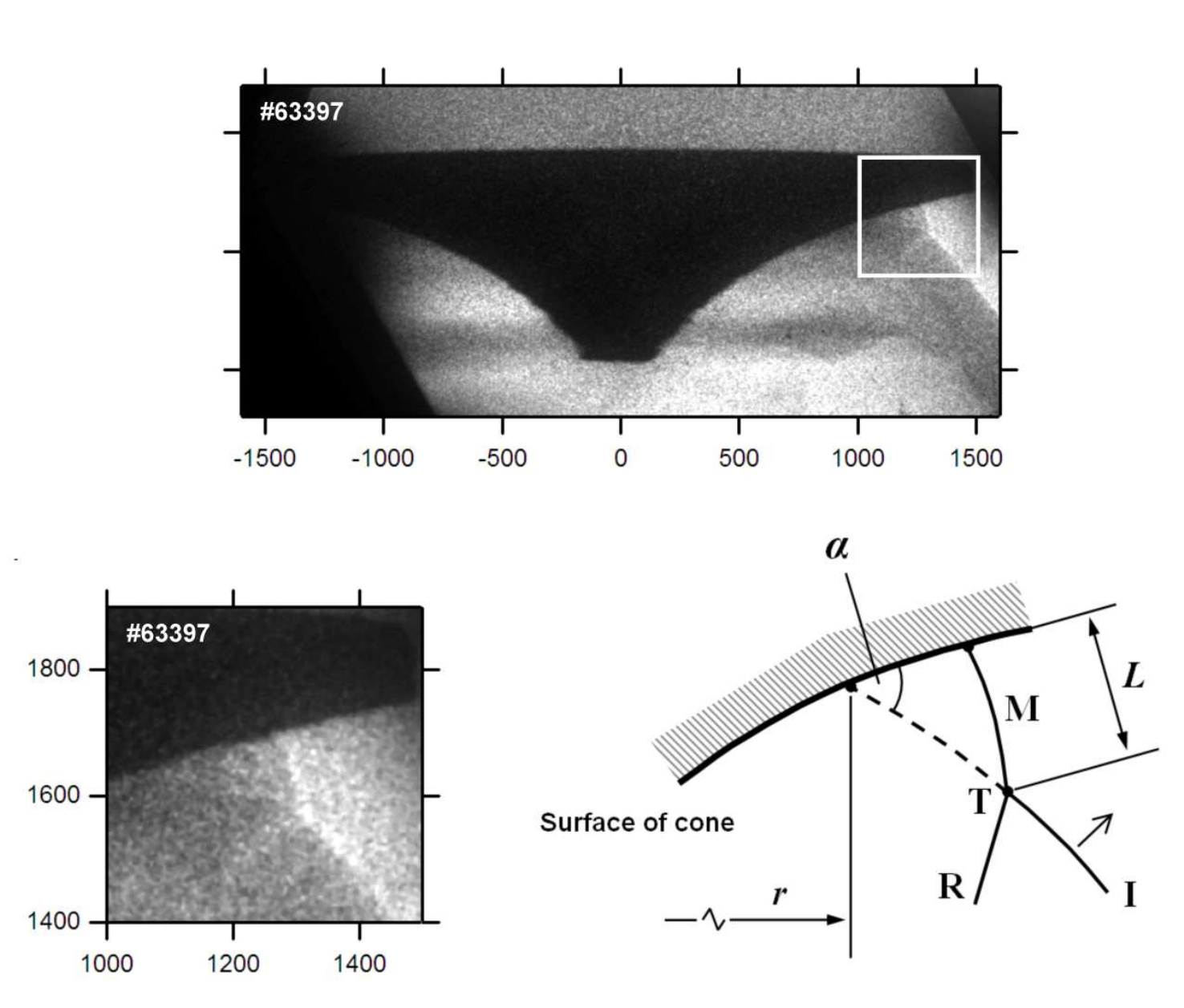}
\caption{
Top: Radiograph of the Mach stem at the surface of the cone. Bottom left:
Enlargement of the boxed area. The spatial scale is in $\mu$m.
Bottom right: Schematic of the radiographs. T is the triple point between
the incident shock (I), reflected shock (R) and Mach stem (M). The
incident shock meets the cone at a radial position $r$ and incident 
angle $\alpha$. The Mach stem size is L.
}
\label{fig:example}
\end{figure}

Figure~\ref{fig:example} shows a typical radiograph from the experiment.
Our goal is to measure how the length (L) of the Mach stem
varies with time. From the experimental data we measure the position and shape of the leading shock wave.
We measure the radial position (r), defined as the distance from the axis of symmetry of the bow shock
to its intersection with the cone directly from the radiographs \citep[see][]{yirak13}.
Likewise, the length of the Mach stem is defined in the images as the distance from the location where the Mach
stem intersects the surface of the cone to the triple point where the Mach stem, bow shock, and
reflected shock meet. We measure the interaction angle $\alpha$ directly from the images.

We attempted several different parameterizations
of the profile z = f(r) of the shock wave, and found that the following functional form
fits the experimental shape very well:

\begin{align}
\label{eqn:fitwave}
\begin{split}
&\theta = ar+br^2,
\\
&{\frac{dz}{dr}} = \tan\theta,
\\
&z = z_\circ + \int_0^r \tan(ar+br^2) dr.
\end{split}
\end{align}

\begin{figure}  %fig6
\centering
\includegraphics[angle=0,scale=1.00]{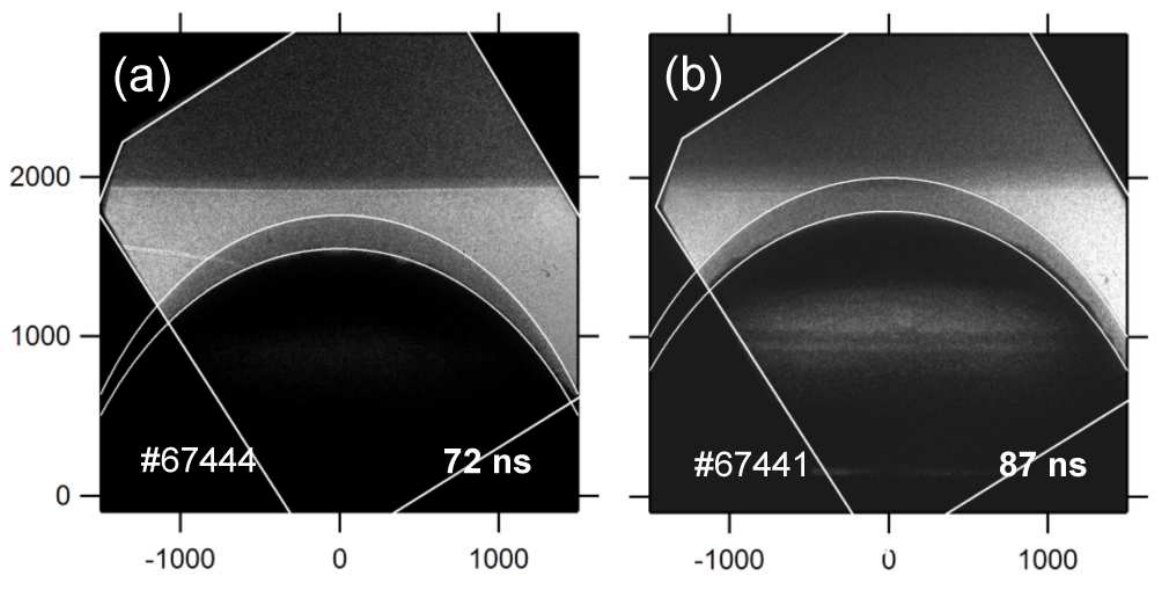}
\caption{
Radiographs (greyscale images) and analytic fits (curved white lines)
to the CH(Br)/shocked-foam
interface (inner curve) and incident shock front
(outer curve) for two experiments. The images were taken at
(a) 72 ns and (b) 87~ns. The bow shock moves from the bottom to
the top in these images. The boxed area is the diagnostic field of view.
Units of the spatial scale are in $\mu$m. The functional form of the analytic
fit is given in the text.
}
\label{fig:waveshape}
\end{figure}

\noindent
The parameters a and b are approximately independent of time, and
z$_\circ$ varies nearly linearly with time. Fig.~\ref{fig:waveshape}
shows a superposition of this parametric fit on experimental
data from targets without an embedded cone.
This parameterization of the shock profile was used to define the
cone profiles used for the experiments summarized in Fig.~\ref{fig:hysteresis}.
For the experiments summarized in Fig.~\ref{fig:cone}, a less-accurate
parabolic functional form was used for the shock front and to mill the shape of
the concave-sided cones. In consequence, and also because of preheat in the
case of these first experiments, the interaction angle is not precisely constant.
These issues are discussed in more detail below.

\section{Experimental Results}

\subsection{Mach Stem Growth Rates and Critical Angles}
\label{sec:critical}

\begin{figure}  %fig7
\centering
\includegraphics[angle=0,scale=0.80]{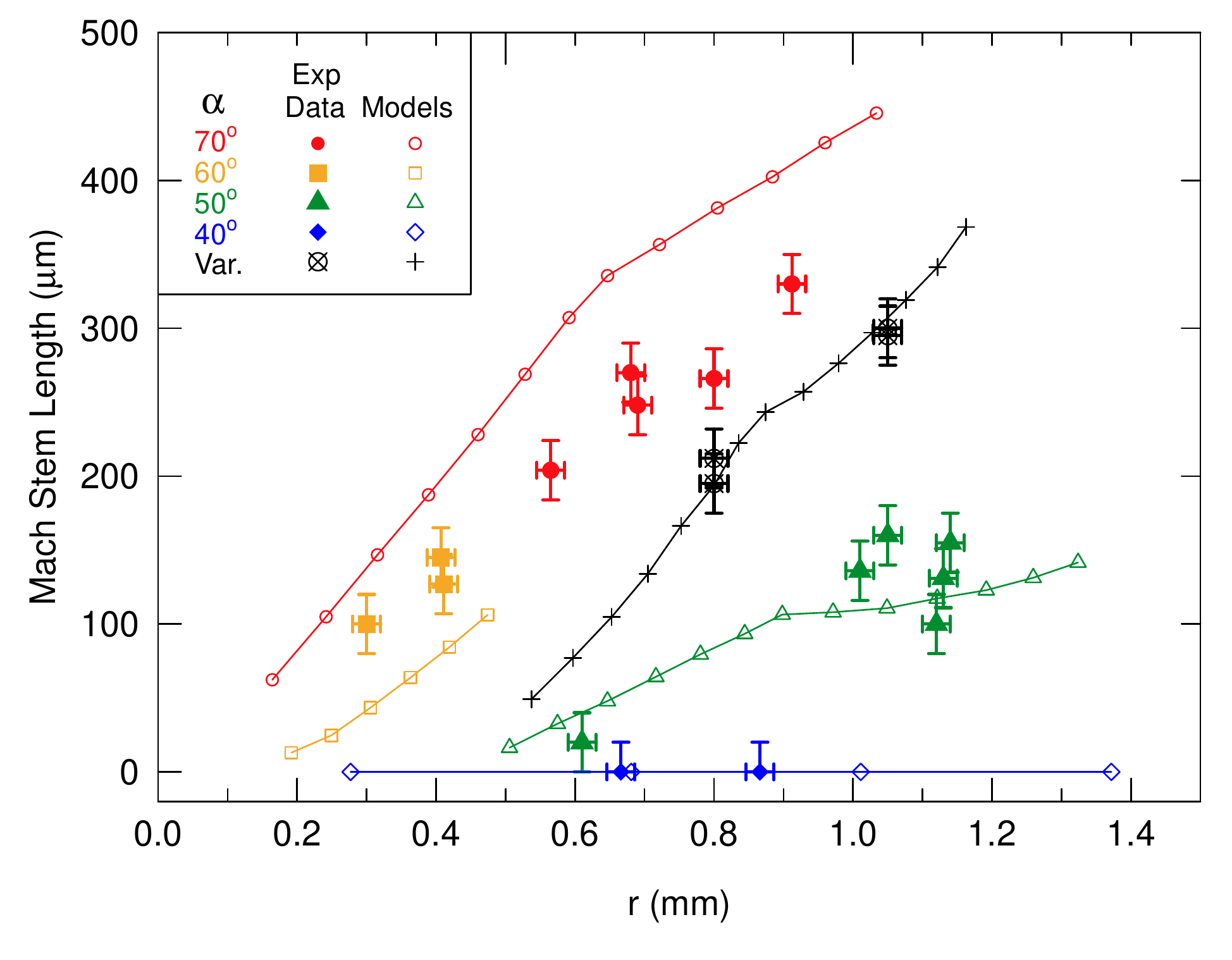}
\caption{Comparison of PETRA simulations with experimental data of Mach stem growth.
Different colors denote curved cones that keep a constant incident angle $\alpha$ 
with the bow shock. Open symbols denote
simulations and filled symbols show experimental measurements. Black points and lines
are for a 40$^\circ$ straight cone that has a variable
incident angle with the curved bow shock
($\alpha$ = 50$^\circ$ at the apex, increasing to $\sim$ 70$^\circ$ at
the edge of the cone).
Errorbars depict measurement uncertainties.
}     
\label{fig:cone}
\end{figure}

Figure~\ref{fig:cone} summarizes the Mach stem growth rates for constant intersection
angles of 40$^{\circ}$, 50$^{\circ}$, 60$^\circ$ and 70$^{\circ}$.
No Mach stems are evident in the 40$^{\circ}$ data, but Mach stems are present in the
50$^{\circ}$, 60$^{\circ}$, and 70$^{\circ}$ plots.  The growth rates
increase from zero at 40$^\circ$ to a substantial fraction of the maximum growth rate
observed by 50$^\circ$, and remain high at $\alpha$ = 60$^\circ$ and $\alpha$ = 70$^\circ$.
The experimental data constrain the critical angle for Mach stem formation $\alpha_C$
to lie between 40$^\circ$ and 50$^\circ$, which implies $\gamma$ = 1.31 $-$ 1.56
using Equation~\ref{eqn:cfeqn}.

The Eulerian radiation-hydrodynamical code PETRA \citep{petra} was used to simulate our experiments.
Eulerian hydrodynamics is essential to treat the large shear flows over
the surface of the gold cones, and the simulations included
the full foil-foam-cone package and the laser-heated hohlraum,
represented on a fixed 1.25 $-$ 2.5 $\mu$m-resolution
mesh.  X-ray diffusion was used to drive the hohlraum evolution and the acceleration
of the ablator-pusher foil. Most of the equations-of-state were taken from the
LANL SESAME library \citep{sesame}.
Mach stem lengths for each numerical model in Fig.~\ref{fig:cone} are connected by lines, and
fit the experimental data reasonably well.  Differences between the models and experimental data
for the 60$^\circ$ and 70$^\circ$ cones likely arise from two complications. First, the 
radiographs from the experiments completed early in the campaign show that the cone
began to ablate before the arrival of the bow shock owing to radiative pre-heating. 
These effects, not included in the numerical simulations,
were reduced in the later experiments by using a pusher designed to be more
opaque to radiation (Sec.~\ref{sec:expr}). The second complication
is that while the curved cones were designed to maintain as constant of an incident angle as
possible, some variation does occur in this angle as the bow shock evolves. In the
numerical simulations we found that $\alpha$ maintained the desired value to within
$\sim$ $\pm$ 4$^\circ$. Variations of this order, combined with the effects of radiative
preheating, are the most likely cause of the differences between models and experiments in
Fig.~\ref{fig:cone}.  Simulations of the potentially more-complex case of a straight cone with a variable
incident angle agree with the experimental data within the uncertainties.

Because the numerical results reproduce the experimental data well, we can use the simulations to
narrow the uncertainties of experimental $\gamma$ considerably.  Models and experimental data
for the dual cones (Sec.~\ref{sec:hysteresis}) show positive growth rates
when $\alpha$ $\gtrsim$ 42$^\circ$ $\pm$ 1$^\circ$. With this range for $\alpha_C$ we find
$\gamma$ = 1.49 $\pm$ 0.03 from Eqn.~\ref{eqn:cfeqn}.  As a check of the
validity of the analytical values,  AstroBEAR models with $\gamma$ = 1.4 of dual bow shocks
described in section~\ref{sec:angles} yield $\alpha_C$ = 43$^\circ$, which, using Eqn.~\ref{eqn:cfeqn},
would imply $\gamma$ = 1.47, a value 5\%\ higher than the actual one.
If we apply this 5\%\ correction to our numerical and experimental value of
$\gamma$ = 1.49, we obtain $\gamma$ = 1.42 for the experimental $\gamma$, with an
observational uncertainty of $\sim$ 2\%, and a systematic uncertainty in translating
$\alpha_C$ to $\gamma$ of $\sim$ 5\%.
The experimental value of $\gamma$ is considerably lower than that
of an ideal monatomic gas (5/3).  Some energy from the shock goes into completing
the vaporization of the foam that was begun by the radiative preheating, and more
energy losses occur from dissociating and ionizing the resultant CH gas.
These processes lower $\gamma$ from the monatomic gas result. 

\subsection{Experiments of Mach Stem Regrowth and Hysteresis}
\label{sec:hysteresis}

Depending in part on whether or not $\alpha$ exceeds the critical value,
Mach stems should either grow, remain stationary with time, or decay.
In the parlance of \cite{SWRF2nd}, these three cases are referred to as
direct, stationary, and inverse Mach reflection, respectively.
Dual-angled cones such as the one shown in panel (b) of Fig.~\ref{fig:dual} have the
desirable property that the incident angle steadily increases, then decreases
sharply to below the critical angle, and finally increases again to greater than
the critical angle, making it possible to study how the Mach stem responds to sudden
changes in the intersection angles.
However, as we investigate numerically in Sec.~\ref{sec:gamma14}, growth rate is not determined
uniquely by $\alpha$ in all cases, but also depends upon the hydrodynamical
flow present in the postshock region.

\begin{figure}  %fig8
\centering
\includegraphics[angle=0,scale=1.00]{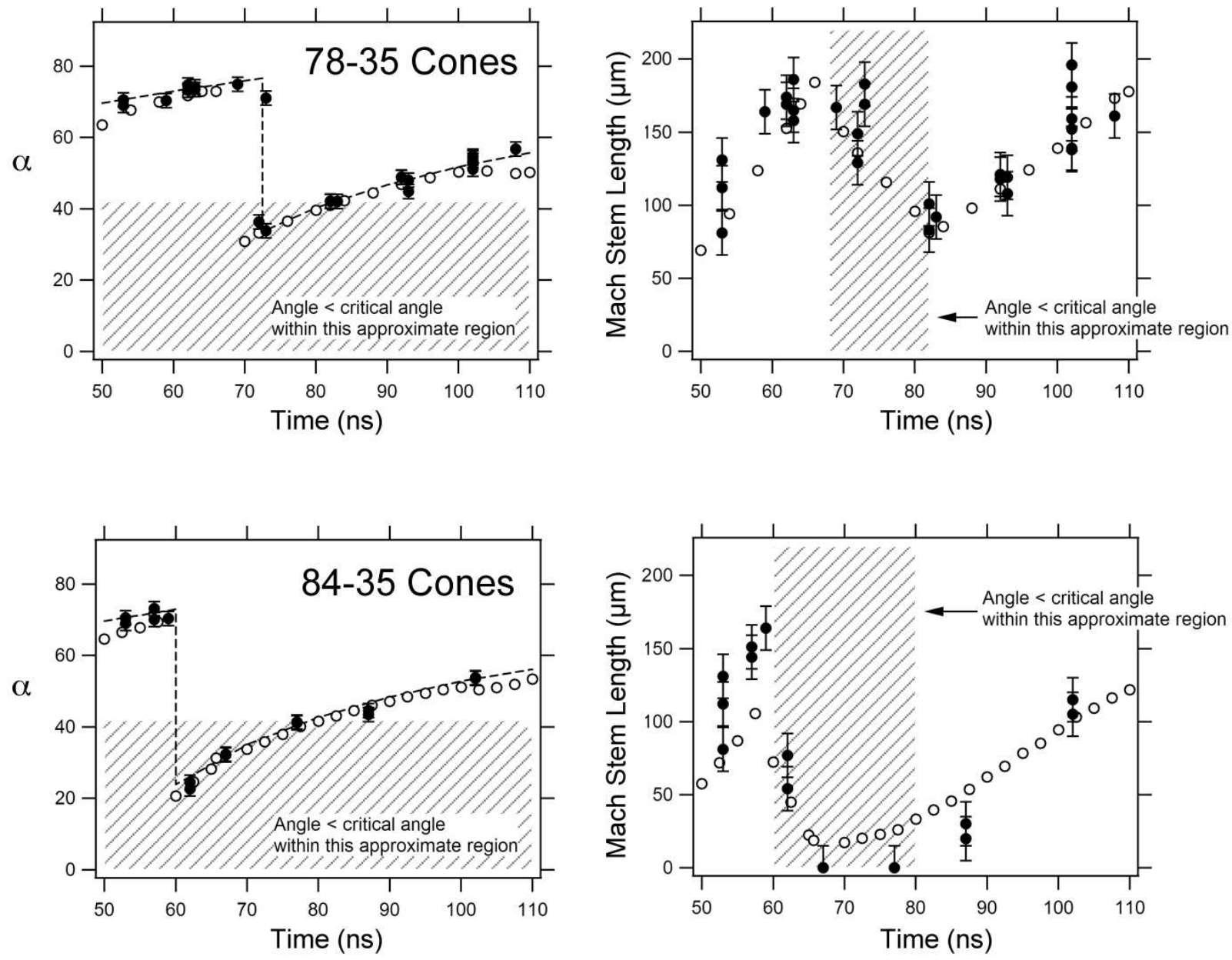}
\caption{
Mach stem destruction and reformation. Top left: The incident angle of the 
shock as a function of time for the dual-angled cone depicted in 
panel (b) of Fig.~\ref{fig:dual}. The incident angle drops 11$^\circ$
below the critical value and then rises above it. Open circles depict values
from AWE PETRA simulations and solid points with errorbars are experimental data.
Dashed-lines are curves fit through the experimental points.
Top right: The Mach stem length
grows or decays depending on if the angle exceeds the critical value of $\sim$ 42$^\circ$.
Bottom: Same as the top panels but for a cone where the angle drops
21$^\circ$ below critical. The Mach stem is destroyed rapidly in this
case, but reforms again once the critical angle is reached.
}
\label{fig:hysteresis}
\end{figure}

Figure~\ref{fig:hysteresis} summarizes results from 
the dual-angled cone experiments. We constructed two types of dual-angled cones,
a set of `78-35' targets where the intersection angle dropped below $\alpha_C$ by $\sim$ 11$^\circ$, and
a set of `84-35' targets that dropped below $\alpha_C$ by $\sim$ 21$^\circ$.
The shaded region in Fig.~\ref{fig:hysteresis} depicts the subcritical angle
regime $\alpha$ $<$ $\alpha_C$ for the two cases. In both types of targets, the
Mach stem decays in size as soon as the intersection angle becomes subcritical.
The decay is rapid for the 84-35 targets and the Mach stem disappears.
However, the Mach stem is not destroyed immediately in the 78-35 targets,
and for both types of targets the Mach stem once again grows as soon as
$\alpha$ $\gtrsim$ 42$^\circ$, which we take to be the critical angle $\alpha_C$.
The top-right panel of Fig.~\ref{fig:hysteresis} shows that the
decay and growth rates on either side of the critical angle are similar.

As noted in the Introduction, Mach stems exhibit hysteresis phenomena 
in the sense that the critical angle of transition from RR to MR differs from
that of MR to RR. Our dual-angled cones cross the critical angle
both from the MR regime into RR, where decay is observed, and from RR into MR, where
growth occurs. However, as these experiments are time-dependent, they do not
test Mach stem formation and destruction as intersection angles are varied in
a quasi-static manner across the critical values, a regime where wind-tunnel
experiments are more optimal. The critical angles in the
targets where $\alpha$ varies rapidly with time are (within the measurement
uncertainties) the same as those inferred from the targets that have constant $\alpha$.
The critical angle for decay is much less-constrained: the data only imply that
11$^\circ$ subcritical suffices to initiate decay, and that the decay occurs more
rapidly at 21$^\circ$ subcritical.

\subsection{Experiments and Models of Mach Stem Survival Along Inhomogeneous Surfaces}
\label{sec:bumpy}

Once a Mach stem grows larger than any surface irregularities, its triple point should
flow over the top of the irregularities. However, the situation when the surface 
irregularities are larger than the triple point is less clear, especially early in
the growth stage when the Mach stem is small. 
When $\alpha$ $<$ $\alpha_C$ we expect the Mach stem to decay, and for this case
\cite{bd87} showed inverse Mach reflection (i.e.~Mach stem decay) proceeds steadily 
and the triple point lowers to the surface. After the triple point contacts the surface,
the system readjusts to a new configuration known as transitioned regular reflection (TRR)
characterized by a leading RR followed by a Mach reflection.

\begin{figure}  %fig9
\centering
\includegraphics[angle=0,scale=1.00]{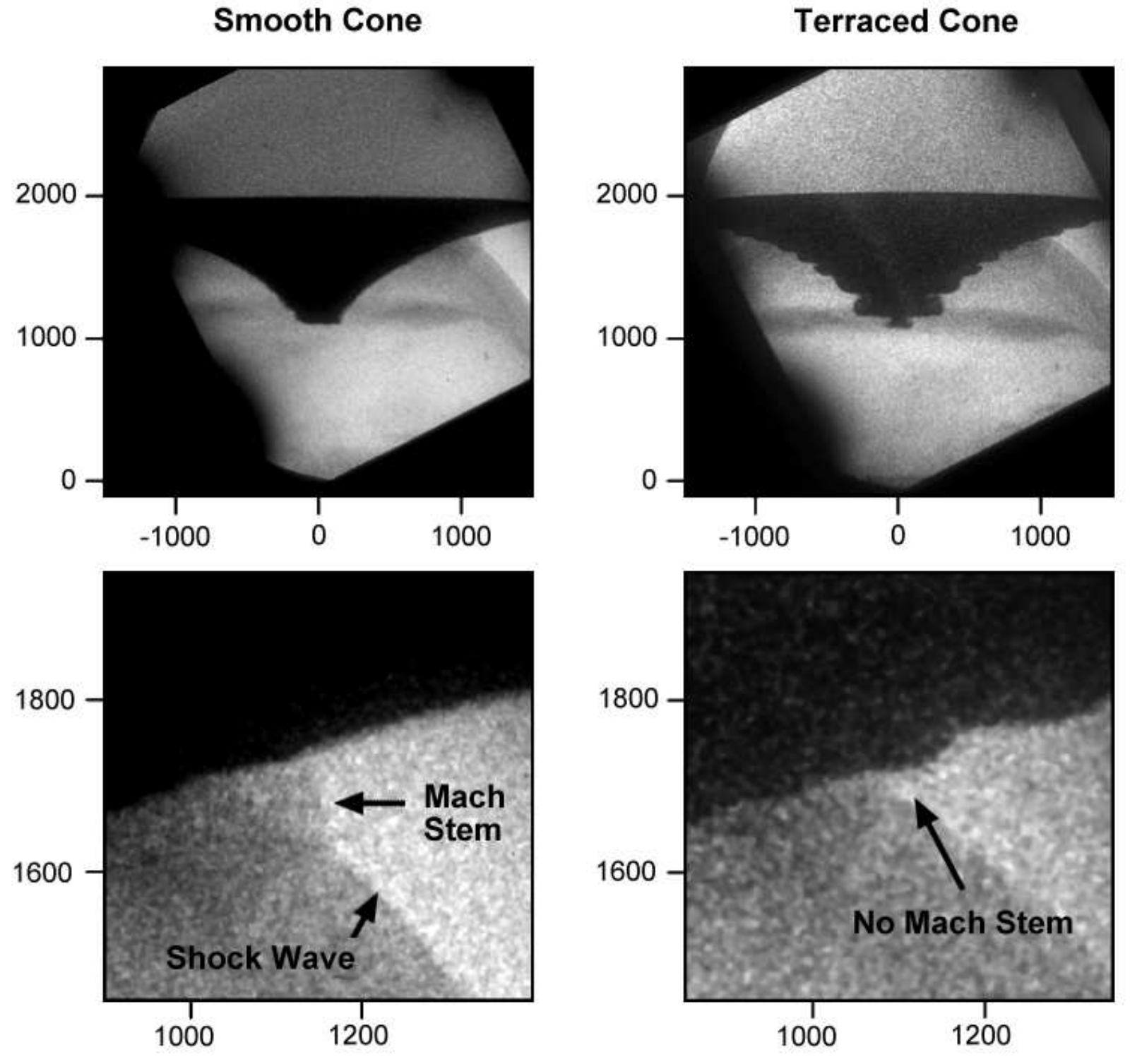}
\caption{Mach stems and irregular surfaces. Left: The radiograph shows 
a well-defined Mach stem where the shock wave meets the surface of the smooth cone.
The laser drives a shock wave from the bottom to the top.
Right: A similar cone with a terraced surface does not produce a Mach stem. Axis
labels are distances in $\mu$m.}
\label{fig:irregular}
\end{figure}

To investigate this case experimentally,
we constructed a terraced cone from a constant-incident angled 50$^\circ$ cone as depicted in
Fig.~\ref{fig:irregular}. As shown in the figure, a prominent Mach stem exists by the
time the bow shock has traversed the smooth 50$^\circ$ cone, while the terraced cone
shows no Mach stem.  The simplest explanation for the lack of a Mach stem
for the terraced cone is that any Mach stem that may form does not grow fast
enough to allow the triple point to clear the next terrace. 
Combined with data in the previous section, this result implies that
although Mach stems can survive sudden and erratic changes in the intersection
angles, they are easiest to create when conditions are relatively smooth.

\begin{figure}  %fig10
\centering
\includegraphics[angle=0,scale=1.00]{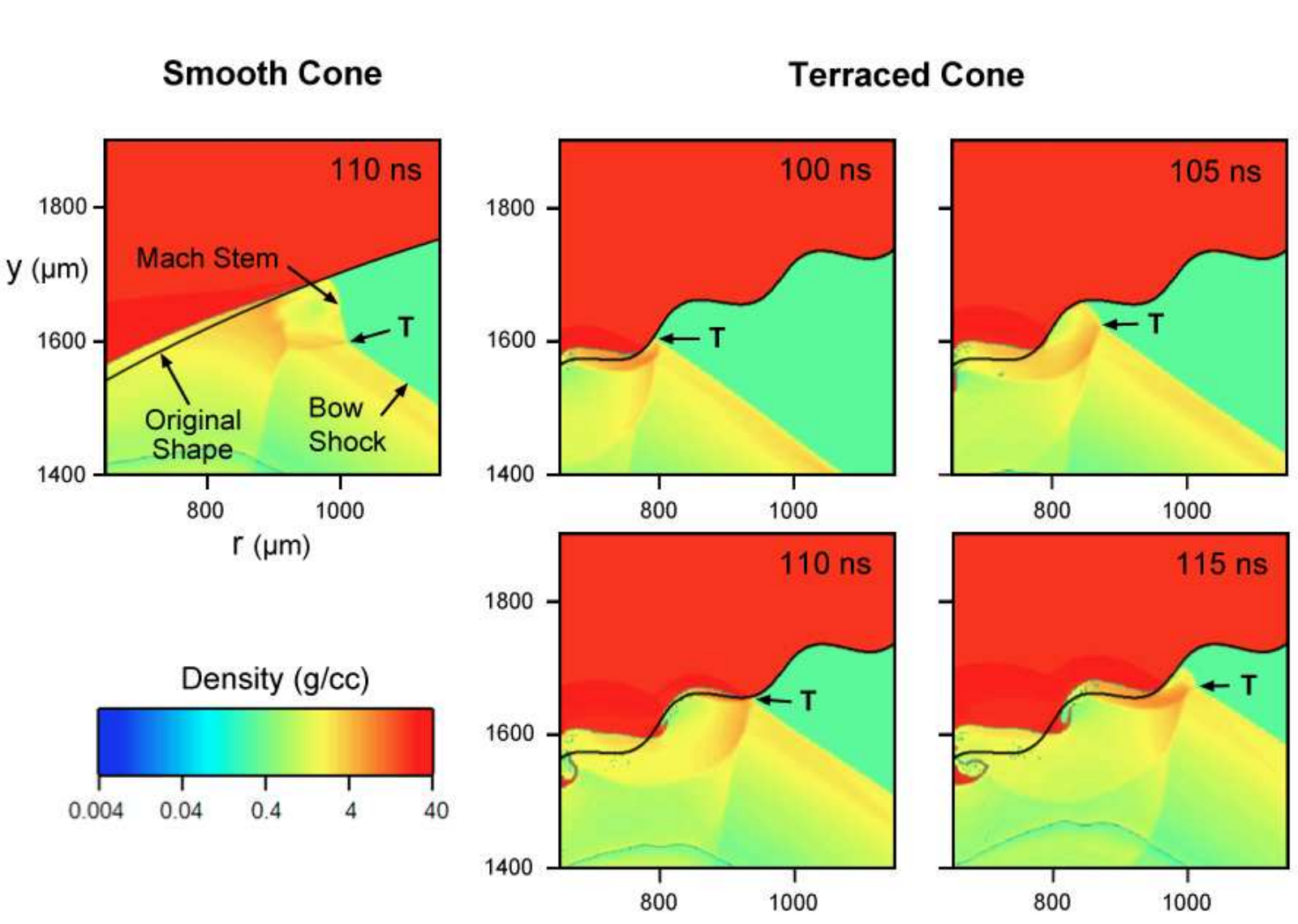}
\caption{PETRA simulations of the bow shock moving along the surface of a smooth cone (left),
and a terraced cone (right). The curved surface of the cone keeps a constant angle of 50$^\circ$
between the bow shock and the cone's surface. The point marked `T' is the triple point. The Mach
stem grows in the valleys and is destroyed in the hills of the terraced cone,
but grows steadily along the smooth cone. Scale sizes are the same as those in
Fig.~\ref{fig:irregular}, and the times in ns after the laser pulse are shown for each panel.
}
\label{fig:bumpysim}
\end{figure}

Numerical models in Fig.~\ref{fig:bumpysim} illustrate this phenomenon. 
After 110 ns a well-defined Mach stem has formed along the surface of the smooth cone.
In contrast, while a Mach stem grows quickly in the valleys of the terraced cone,
the triple point (labeled `T' in Fig.~\ref{fig:bumpysim}) runs into the top of
the next ridge as the bow shock proceeds. Hence, the surface irregularities are large
enough to inhibit Mach stem growth in this case. We see no clear evidence for TRR,
but the situation is far from steady-state and the spatial resolution of the
simulations may need to be higher to resolve this feature.

\section{Physics of Mach Stem Evolution and its Astrophysical Implications}

Why do Mach stems form?  As described in the Introduction
and depicted in Fig.~\ref{fig:MS-cartoon}, a triple point is sometimes needed
to satisfy the boundary conditions when two shocks intersect. 
Interpretations that involve more physical insight are
discussed in detail by \cite{SWRF2nd}.
One school of thought is that the
critical angle $\alpha_C$ for Mach stems
is set when the gas behind the reflected shock becomes subsonic
\citep[e.g.][]{vn43}. Another point of view is that
the critical angle occurs when the compression behind the reflected shock matches
that of a normal bow shock, which would allow a Mach stem to grow without a sudden pressure
change \citep{hl75}.  \citet{SWRF2nd} remarks that the supersonic criterion 
appears to be the one best supported by experimental data. 

\subsection{Minimum and Maximum Critical Angles for Intersecting Bow Shocks}
\label{sec:angles}

We conducted a series of numerical simulations in order to clarify how 
Mach stems form, evolve, and dissipate when they result from intersecting
bow shocks. The simulations were done with the AstroBEAR code in 2-D
(planar symmetry).  AstroBEAR is a fully 3-D MHD parallelized
code with adaptive mesh and cooling capabilities \citep{ab1,ab2}.
The setup used gases with fixed $\gamma$ = 1.67, 1.4 and 1.2,
and Mach numbers M = 1.6, 5.2, and 30. These parameters are shown alongside
the experimental and observational values in Table~1.

In the numerical models, two
circular obstacles with diameters of d$_\circ$ = $3\times10^{14}$~cm
were embedded into a 50~km$\,$s$^{-1}$ wind, and for each of the nine combinations
of $\gamma$ and M, the simulation followed the evolution for
$\sim$ 40 $\tau_F$, where the flow time $\tau_F$ = 1.9~years
is defined as the time it takes unperturbed wind to move one obstacle diameter.
The density of the wind was $5\times 10^3$ cm$^{-3}$, and the
initial density contrast was $2\times 10^3$ between the obstacles and the incident
flow. We also investigated a model with a higher-density contrast of $2\times 10^5$ 
to assess how much deformations in the obstacles affected the results.

\subsubsection{Numerical Simulations for $\gamma$ = 1.4 and M = 5.2}
\label{sec:gamma14}

%t = 0 is at frame 7
%Each frame = 0.75 years = 0.3947 t_f; where t_f = 1.9 years

\begin{figure}  %fig11
\centering
\includegraphics[angle=0,scale=0.975]{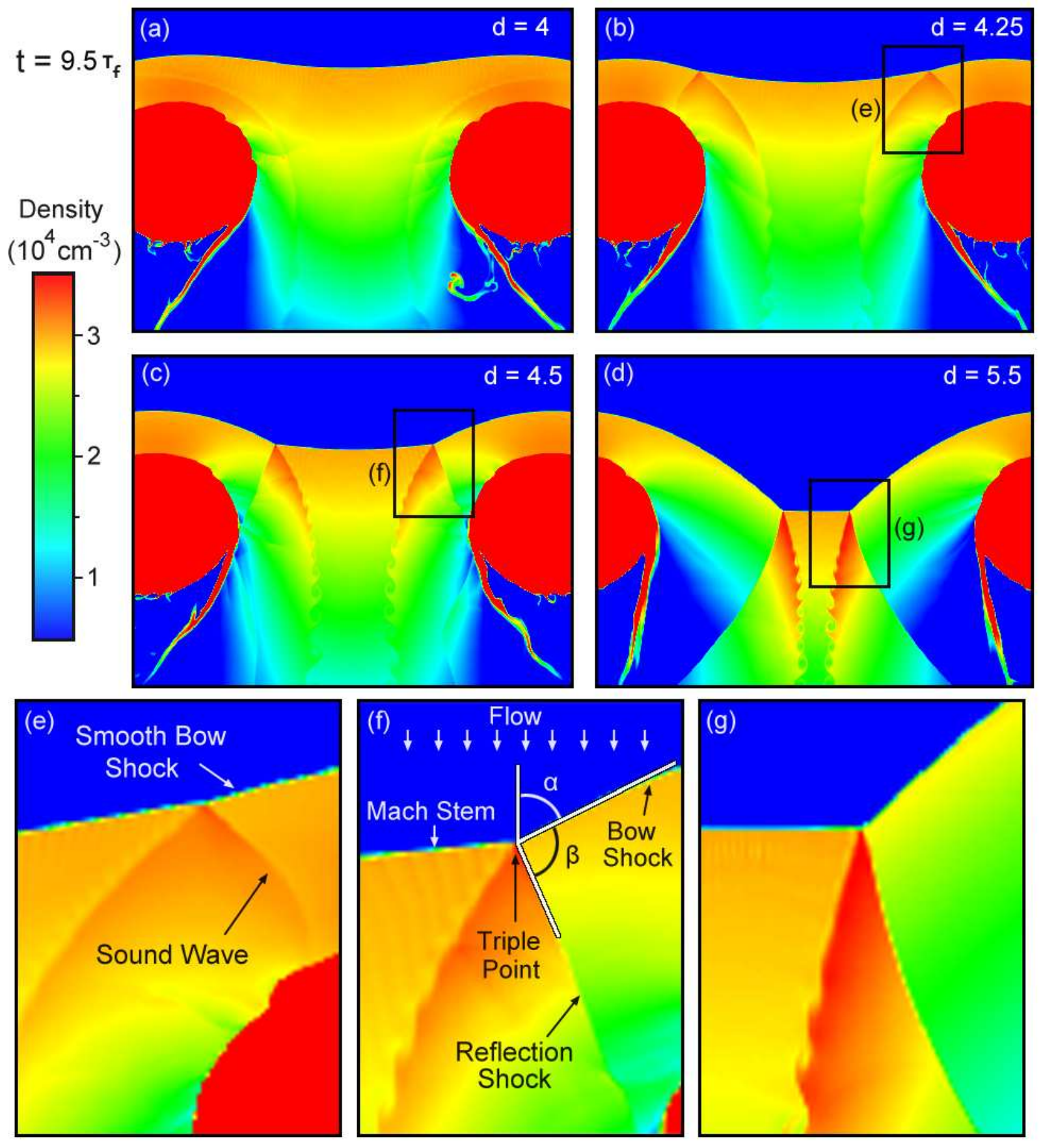}
\caption{Top and Middle: AstroBEAR 2-D simulations for $\gamma$ = 1.4 and M = 5.2
of intersecting bow shocks at t = 9.6$\tau_F$, where $\tau_F$ is the time it takes for
the unperturbed flow to traverse the diameter of the obstacle. The colors denote different densities.
The parameter d refers to the initial separation between the obstacles in units of the obstacle radius. Bottom:
Expansion of the boxed regions around the triple points. The angles $\alpha$ and $\beta$
are discussed in the text.
}
\label{fig:sim1}
\end{figure}

\begin{figure}  %fig12
\centering
\includegraphics[angle=0,scale=1.00]{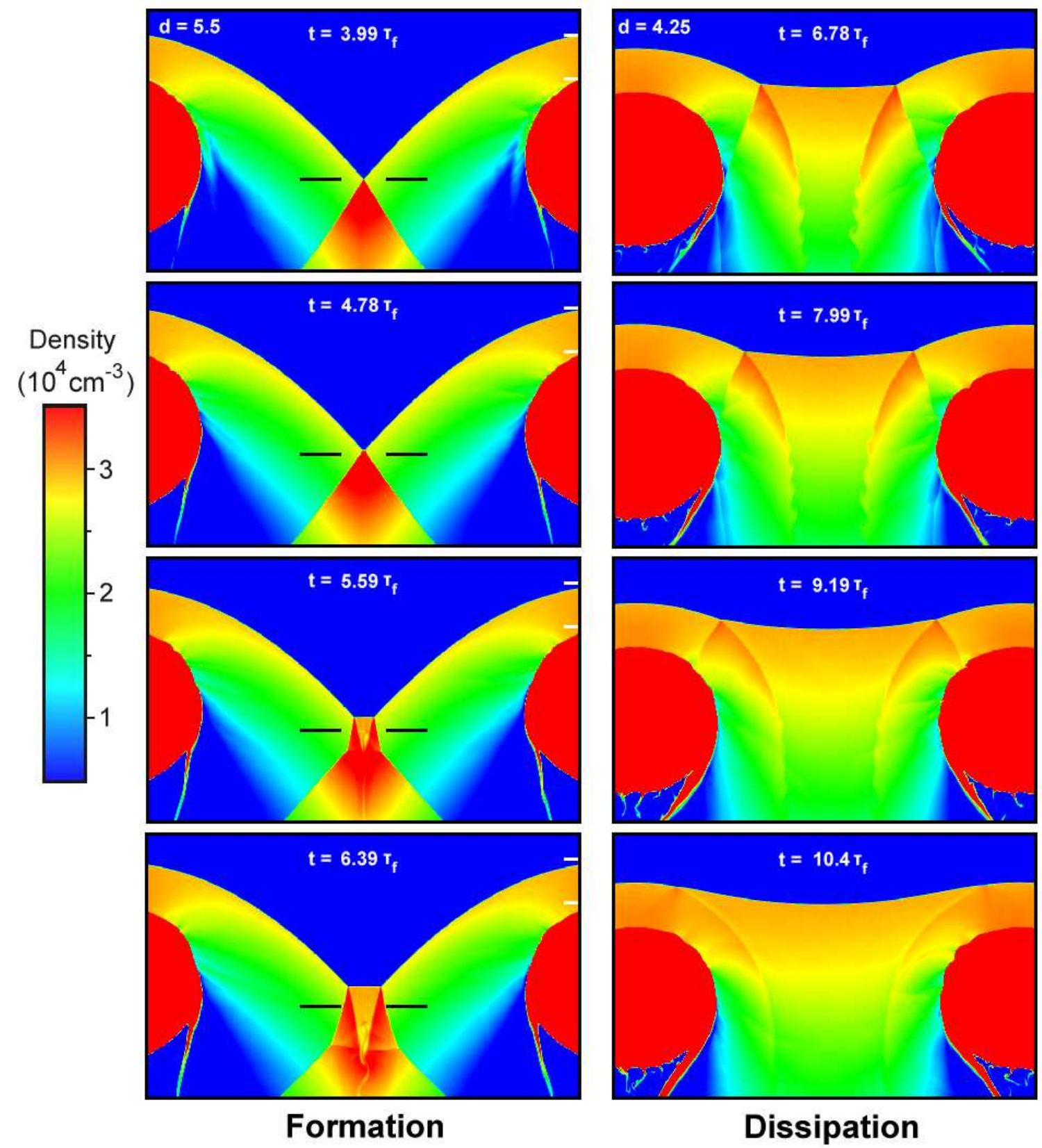}
\caption{Left: Density plots of the time-evolution of the triple point
in the d = 5.5 r$_\circ$ case shown in  Fig.~\ref{fig:sim1}, which begins as
regular reflection and develops a Mach stem. Black fiducials mark the initial location of the
triple point, and white fiducials mark the initial locations of the bow shock and
obstacle boundary. Right: Time-sequence of the triple point
for d = 4.25 r$_\circ$.  As the system approaches, and then exceeds $\alpha_M$, the 
reflection shocks become sound waves, the triple point dissipates, and the bow shock
becomes smooth.  The color scale is the same as those in Fig.~\ref{fig:sim1}.
}
\label{fig:sim2}
\end{figure}

The $\gamma$ = 1.4 simulations with M = 5.2 are summarized in Figs.~\ref{fig:sim1} and~\ref{fig:sim2}.
At any given time, the simulations exhibit one of three configurations
determined primarily by the separation between the obstacles and
the amount of time the system has had to evolve: (i) regular reflection
where the bow shocks meet, 
(ii) a Mach stem between the bows, or (iii) a single smooth shock that
encompasses both obstacles. For the $\gamma$ = 1.4 models,
when the obstacle separation d $\gtrsim$ 6 r$_\circ$
(r$_\circ$ is the obstacle radius) the intersection angle $\alpha$ between the
bow shocks is below the critical angle $\alpha_C$, and the system shows
regular reflection for the length of the simulation. 
If the obstacles are close enough that $\alpha$ is not too far below $\alpha_C$
(e.g. Model d = 5.5 r$_\circ$ in Fig.~\ref{fig:sim2}), the incident angle
$\alpha$ may grow enough with time to exceed $\alpha_C$ at some point, which causes
a Mach stem to form after an initial period of regular reflection 
(left panels of Fig.~\ref{fig:sim2}). Both the d = 5.5 r$_\circ$ and d = 5.0 r$_\circ$ models
settle into a quasi-steady-state configuration with a stable Mach stem
at later times. In addition to the triple point where the Mach stem, bow
shock and reflection shock meet, there is also a contact discontinuity that marks the
boundary between gas that enters the Mach stem and gas that goes through
the bow shock and then the reflection shock. This contact discontinuity shows
a significant amount of Kelvin-Helmholtz growth in the simulations.

Mach stems that occur at smaller separations (d $\lesssim$ 4.5 r$_\circ$) grow steadily
with time as the location of the intersection point between the bow shocks moves upstream,
raising $\alpha$.  As $\alpha$ approaches 90$^\circ$ and the triple point
moves closer to the apices of the bow shocks, the
bow shocks redirect postshock material towards the triple point as they do 
when $\alpha$ is smaller, but at a lower velocity relative to the triple point.
Once the motion of the postshock gas relative to the
reflection shock becomes subsonic, the reflection shock becomes a sound wave, and the 
numerical simulations show that the Mach stem merges with the bow shock to
produce a single curved shock around the obstacles.
This behavior is shown in the right panels of Fig.~\ref{fig:sim2}.
Hence, there is a maximum angle $\alpha_M$ above which the triple point
and the Mach stem disappear and are replaced by a smoothly-varying bow shape. 

\begin{figure}  %fig13
\centering
\includegraphics[angle=0,scale=0.99]{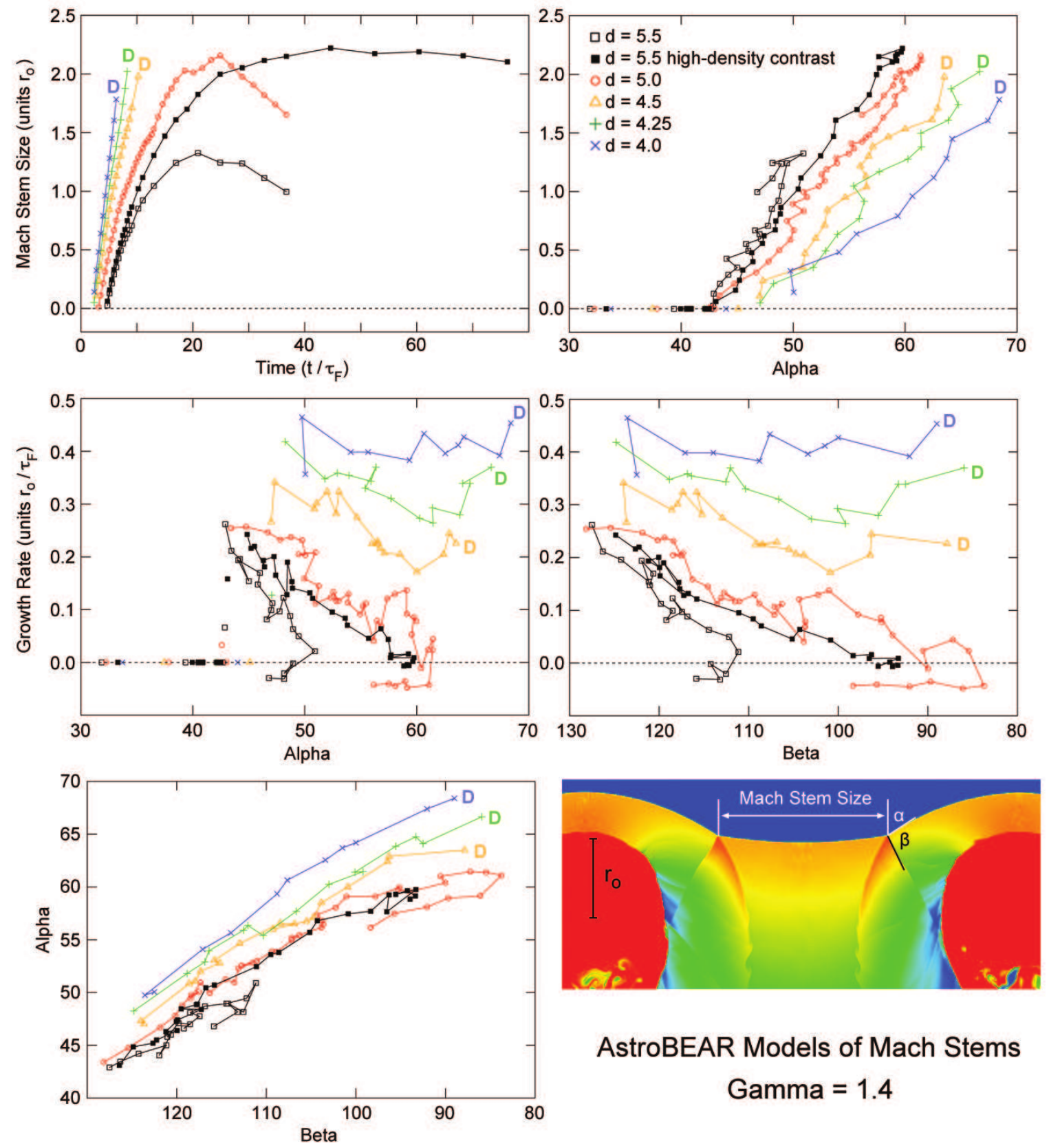}
\caption{AstroBEAR models of intersecting bow shocks for $\gamma$ = 1.4 and M = 5.2.
Top: Mach stem size as a function of time and of the incident angle $\alpha$ defined in the lower right panel.
The radius of the obstacles is r$_\circ$ and the time for the preshock gas to flow past the
obstacle is $\tau_F$. The largest Mach stems have size $\sim$ 2 r$_\circ$.
The symbol `D' indicates where the Mach stem dissipated as it became subsonic.
Middle: Mach stem growth rates as a function of the angles $\alpha$ and $\beta$. Growth rates that
drop to zero imply stable Mach stems. Bottom: Relationship of $\alpha$ to $\beta$ in the models.
}
\label{fig:sim3}
\end{figure}

Our numerical models indicate that Mach stems only survive as long
as the angle $\beta$ between the bow shock and the
regular reflection shock at the triple point exceeds $\sim$ 90$^\circ$
(Fig.~\ref{fig:sim3}). The maximum diameter of a Mach stem in the
simulations is $\sim$ 2 r$_\circ$.  For these transient Mach stems, as
$\alpha$ increases, $\beta$ decreases, and when $\beta$ falls below $\sim$
90$^\circ$ the lateral motion into the reflection shock becomes subsonic, and the
wave detaches from the bow, which is now a smoothly-varying curved surface.
Models with smaller separations exhibit higher growth rates and
evolve into a single bow shock more quickly (Figs.~\ref{fig:sim1} and \ref{fig:sim3}).

The value of $\alpha_M$ ranges between 62$^\circ$ and 68$^\circ$ in our $\gamma$ = 1.4,
M = 5.2 simulations for the different values of initial obstacle separation.
The precise value depends upon how much the Mach stem is curved, which
alters the angles at the triple point. Mach stems with more widely-separated obstacles
curve more, and the curvature is also influenced by
how rapidly the flow destroys the obstacles. In all cases,
detachment of the triple point from the bow shock begins when $\beta$ $\lesssim$ 90$^\circ$.
Hence, transient Mach stems form and grow over a range of $\alpha$ $\sim$ 20 $-$ 25 degrees,
and eventually result in a single smooth shock on timescales that range from a few obstacle crossing
times, to an order of magnitude higher than this value.
In the d = 5.5 r$_\circ$ and d = 5 r$_\circ$ models, the Mach stem remained in
quasi-static equilibrium for the entire length of the simulation ($\sim$ 40 crossing times).

The high-density-contrast model with d = 5.5 r$_\circ$ (filled black squares in Fig.~\ref{fig:sim3})
resembles that of the lower-density contrast models d = 5.5 r$_\circ$ at early times, and
d = 5.0 r$_\circ$ at later times. Obstacles in the
lower-density contrast model are more affected by compression and undergo more
mass stripping from the wind. However, the effects of changing the density contrast
are minor compared with changing the separation, evolution time, $\gamma$, and Mach 
number (see below).

\subsubsection{Numerical simulations for other values of $\gamma$ and the Mach number M}
\label{sec:other-gamma}

Our simulations with different values of $\gamma$ and M follow the same pattern
as those for $\gamma$ = 1.4 and M = 5.2 described above. At early times, systems
where the object separations are small show individual bow shocks and a Mach stem. 
At later times, these shocks grow to form a single curved shock
that encompasses both obstacles.
No Mach stems form at when the obstacle separations are large, while stable
Mach stems exist at intermediate separations.

We measured the critical angle $\alpha_C$ for Mach stem formation in each combination
of $\gamma$ and M, and compile these results in Table~2. The last columns in the Table show the 
predicted values of $\alpha_C$ in the limit of infinite M for both the \citet{cf48}
formalism (Equation~\ref{eqn:cfeqn}) and from Equation~6 of \citet{derosa92}.
Both analytic equations explain the simulation results reasonably well for $\gamma$ =
1.67 and 1.4, but the \citet{derosa92} equation matches much better at $\gamma$ = 1.2.
The critical angle increases as the Mach number approaches unity for all values
of $\gamma$.

\begin{figure}  %fig14
\centering
\includegraphics[angle=0,scale=0.80]{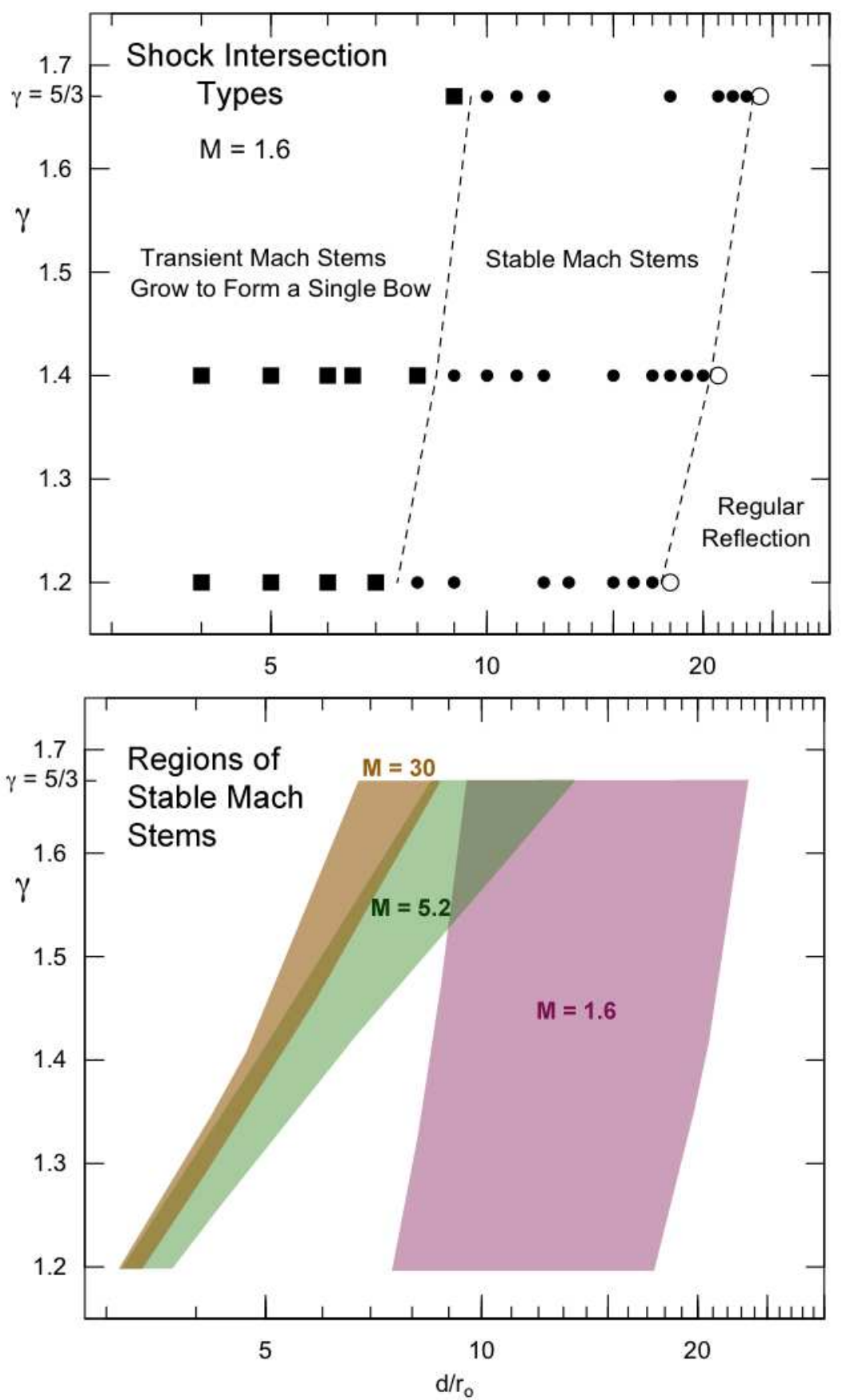}
\caption{Top: Mach stem behavior after 40 dynamical timescales for 2-D AstroBEAR
simulations with Mach number = 1.6, as a function of the polytropic
index $\gamma$ and the logarithm of the separation d between two cylindrical obstacles 
of radius r$_\circ$. Squares are transient Mach stems, filled circles are stable Mach stems,
and open circles are regular reflections.
Bottom: Ranges of stable Mach stems as a function of the polytropic index
$\gamma$, the obstacle separation d/r$_\circ$, and the Mach number M. Stable Mach stems form
most readily with larger $\gamma$ and smaller M.
}
\label{fig:mssummary}
\end{figure}

The range of separations that allow stable Mach stems to form
depends upon both $\gamma$ and M.  The top panel of Fig.~\ref{fig:mssummary}
shows that stable Mach stems exist for larger obstacle separations
and persist over a wider range of distances when $\gamma$ is larger.
For example, when M = 1.6, stable Mach stems exist after 40 $\tau_F$ when
7.5 $\lesssim$ d/r$_\circ$ $\lesssim$ 17.5 for $\gamma$ = 1.2, 
8.5 $\lesssim$ d/r$_\circ$ $\lesssim$ 20.5 for $\gamma$ = 1.4, and
9.5 $\lesssim$ d/r$_\circ$ $\lesssim$ 23.5 for $\gamma$ = 5/3. 
At smaller separations, transient
Mach stems grow to encompass both obstacles, and at larger separations
the bow shock wings intersect with regular reflection.
In all cases, the maximum Mach stem diameter is $\sim$ 2 r$_\circ$.
The bottom panel of Fig.~\ref{fig:mssummary} shows that increasing the Mach
number narrows the range of distances that produce stable Mach stems, and
also moves this range to smaller separations.

We can understand the dependence of stable Mach stems on $\gamma$ and M by
considering the dynamics of the postshock gas.
As $\gamma$ decreases, the postshock pressure declines and bow shocks wrap more
closely around obstacles. For the case of two obstacles considered here,
it means that the obstacles must lie closer together for the intersection
angle to exceed the critical value to form a Mach stem. 
Similarly, when the preshock temperature
declines, increasing M, the postshock pressure is less important dynamically and
bow shocks also wrap more closely around the obstacles. Both effects appear in
the bottom panel of Fig.~\ref{fig:mssummary}.  The effect of the postshock pressure on the shape
of bow shocks has been known for several decades,
for example, in numerical simulations of cooling jets where a
dense plug forms around the jet at the leading working surface \citep{blondin90}.  The most
extreme case is that of an isothermal shock ($\gamma$ = 1), where cooling to
preshock temperatures occurs immediately behind the shock, and no Mach stems form. 

\subsection{Possible Astrophysical Examples}

\begin{figure} %fig15
\centering
\includegraphics[angle=0,scale=1.00]{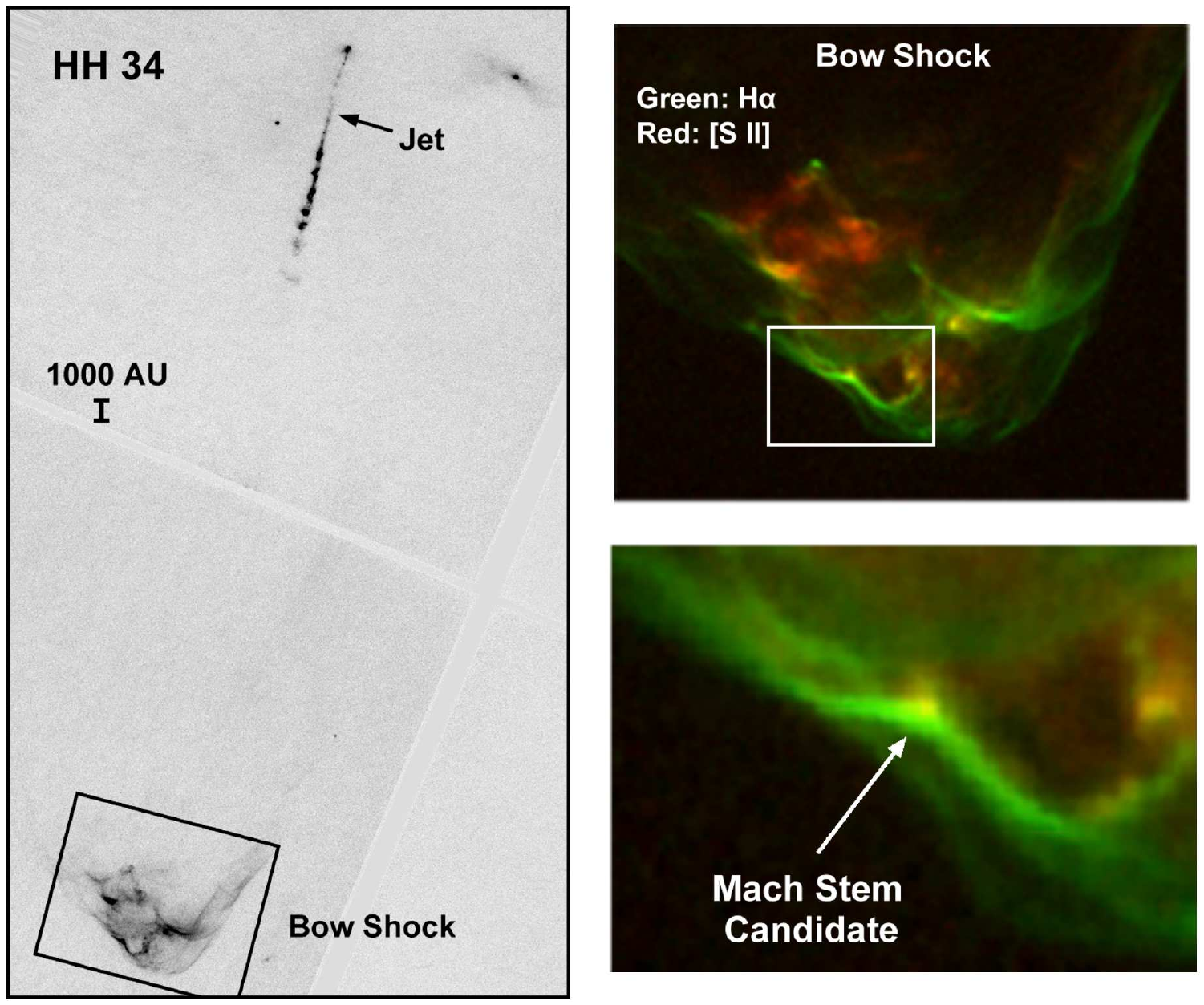}
\caption{Candidate Mach stem within an astronomical outflow. Left: HST image of the
young stellar jet HH34 and its bow shock (from \citep{hartigan11}). Right: Expanded views of
the bow shock and the putative Mach stem. 
}
\label{fig:hh34machstem}
\end{figure}

From an astrophysical standpoint, Mach stems are potentially important because they
change the geometry of shock intersections from oblique to normal, and 
can produce a quasi-static or transient `hot spot' of enhanced emission and excitation at this location.
A prime candidate for this phenomenon exists in the bow shock of the HH 34 stellar jet
\citep{hartigan11}.  Figure~\ref{fig:hh34machstem} shows that a bright emission knot
exists immediately behind the intersection point of two arc-shaped shocks. This knot,
as well as several others in this object, have anomalous proper motions that do not
follow along with the bulk motion of the bow shock. Anomalous pattern motion is expected
if the proper motions simply trace the location of the shock intersection points.

The structured bow shock in HH~47 \citep{hartigan11} is another potential example
where transient Mach stems may occur. This
object resolves into what appear to be dozens of small knots in high-resolution images. 
Surprisingly, these bright points form and dissipate on timescales of a decade, 
which is difficult to explain if they represent discrete knots of high density that
plow through the working surface of the flow. In
both HH~34 and HH~47, the shock velocity is $\sim$ 100 km$\,$s$^{-1}$ and the knot size
$\sim$ 7.5$\times 10^{14}$ cm, so $\tau_F$ $\sim$ 2.4 years. The numerical models
imply we should expect any transient Mach stems to evolve on a timescale of 
about a decade, in agreement with the observations. 

The postshock regions of cooling astrophysical shocks are complex, and contain various
zones of ionization for each element, so these regions are not characterized by a single value of $\gamma$.
However, the overall dynamical effects of cooling in bow shocks has been well-established
for some time \citep[e.g.][]{blondin90}.  A critical parameter
is the ratio of the cooling-zone distance d$_{COOL}$
to the obstacle size d$_\circ$. When this ratio is $\gtrsim$ 1, the gas acts like
an adiabatic flow with $\gamma$ = 5/3, where the maximum size of a Mach stem will be comparable to
d$_\circ$. In the opposite limit when the obstacle is very
large compared with the cooling length, the gas behaves dynamically as if
it is isothermal.  An isothermal shock has $\gamma$ = 1, where no Mach stems are possible.
In this case, Mach stems only exist on scale sizes $\lesssim$ d$_{COOL}$. Hence, the maximum size
of a Mach stem should be the smaller of d$_{COOL}$ or d$_\circ$.
The cooling distance scales inversely as the preshock
density and directly as a power of the shock velocity \citep[e.g.][]{hrh87}, and also depends
upon the strength of the magnetic field \citep{hw15}.
Hence, whether or not two intersecting bow shocks create a substantial Mach stem depends upon the
shock velocity, density, magnetic field strength, as well as the size of the bows.
Overall, Mach stems should be most common when size scales are small enough to render
cooling unimportant in the dynamics.

The distribution of clumps in the flow and ambient medium ultimately determines whether or not
Mach stems will form and grow in any situation, with the optimal separation for Mach stem
growth being several times the diameter of the obstacles in a flow with moderate Mach number.
The widest range of allowable obstacle separations occurs when cooling is least important
(Sec.~\ref{sec:other-gamma}).
Cooling distances in the HH~34 and HH~47 bow shocks are on the order of the
size of the transient bright knots in these objects \citep{hartigan11}. Hence, the effective $\gamma$
for these spatial scales is $\sim$ 5/3, favorable for Mach stem growth.
We present simulations of intersecting shocks in 3-D, driven by velocity-variable jets with
accurate radiative emission-line cooling in a future paper (Hansen et~al., 2016, submitted to ApJ).

\section{Summary}

In this paper we examined hydrodynamical phenomena associated with Mach stems, which
are normal shocks that occur whenever two shock waves intersect one another
within a certain range of angles.  Using the Omega laser facility
we created Mach stems in high-Mach number plasmas under controlled
conditions in the laboratory with cone-shaped targets,
and complemented this work with numerical simulations of the
experiment and for the more astrophysically-relevant case of intersecting bow shocks.
Our main focus has been to understand how Mach stems respond as the angle between
the two incident shocks (equivalently, the angle between the shock wave and a surface)
varies in supersonic plasmas.

Our first set of experiments employed a design that enforced a constant angle
between a curved shock and a conical surface. These experiments demonstrated that the
Mach stem growth rate was highest for the largest incident angles and the rate 
increased most rapidly when the intersection angle was closer to the critical value
$\alpha_C$.  The measured value of $\alpha_C$ from the constant-angle cones was 
between 40$^\circ$ and 50$^\circ$. Experiments with dual-angled cones allowed for a more precise
estimate of $\alpha_C$, as well as a means to quantify Mach stem decay rates
when $\alpha$ $<$ $\alpha_C$. After using the experimental growth rate to 
verify simulations of the experiment, we combined the simulations and experimental data
together to refine the critical angle $\alpha_C$ in the experiments to
42$^\circ$ $\pm$ 1.0$^\circ$. Numerical simulations with different values of
$\gamma$ and the Mach number M made it possible to test analytical formulae
that translate $\alpha_C$ to $\gamma$,
and we found that the relatively simple formula from \cite{cf48} reproduces the
numerical results with approximately a 5\%\ error as long as $\gamma$ $\gtrsim$ 1.4. 
At smaller values of $\gamma$, only the more complex
analytic formula for $\alpha_C$ from \citet{derosa92} is consistent with the simulations.
Our best estimate for the effective $\gamma$ of the shocked foam is $\sim$ 1.42.  
This value differs significantly
from the ideal monatomic gas $\gamma$ of 5/3 owing to the 
energy lost to dissociation and ionization of the gas
once the solid has been vaporized. 

Simple geometrical considerations argue that Mach stems should dissipate between
intersecting bow shocks if the stems grow too large, and our numerical simulations confirm this hypothesis. 
When the angle between the bow shock and the outer reflection shock
becomes acute, the reflection shocks at the triple point become subsonic, and the
triple point dissipates as the Mach stem joins smoothly with the main bow shocks. The critical
incident angle at which this occurs depends upon the curvature of the Mach stem,
and ranges between 62$^\circ$ and 68$^\circ$ for $\gamma$ = 1.4. Hence, there is
a range of incident angles of $\sim$ 25 degrees over which Mach stems occur.

Bow shocks that intersect produce one of four outcomes: (i) steady-state regular reflection,
(ii) transient regular reflection that evolves into a Mach stem, 
(iii) a steady-state Mach stem, or (iv) a transient Mach stem that grows to envelop
both obstacles and produce a single smooth shock at late times. As the separation between the
obstacles decreases, systems transition from cases (i) and (ii) to cases (iii) and (iv). The
range of separations that produce Mach stems is higher for larger $\gamma$ and lower Mach
numbers. The maximum
size for a Mach stem is comparable to the diameter of the obstacles, and the characteristic
lifetimes for most transient Mach stems are $\sim$ 10 flow times. 

Our experiments show that Mach stems persist in high-Mach number plasma shocks, and are
relatively robust in the sense that when the intersection angle drops below
the critical value the Mach stems decay but are not destroyed immediately.
The decay rate in the size of the Mach stem at subcritical angles appears
comparable to the growth rates for supercritical angles. In our experiments,
after a sudden decrease of the intersection to subcritical, the
angle gradually rose until it once again exceeded the critical value, after which the
Mach stem again grew. 
In real astrophysical situations the ambient medium can be highly nonuniform, 
resulting in bumpy shock fronts. Our experiments show that a
rough surface can inhibit Mach stem growth if the Mach stem does not grow fast
enough to flow past the surface irregularities. Simulations with $\gamma$ = 1.2 and
$\gamma$ = 5/3 show that cooling also significantly inhibits Mach stem growth, so that Mach stems
are most likely to form on size scales $\lesssim$ d$_{COOL}$.

Despite these restrictions, there is some observational evidence for 
Mach stems in an astrophysical context. Mach stems are an attractive explanation
for the bright knots in the HH~34 bow shock,
both from the standpoint of morphology and from anomalous pattern motions that
appear to trace intersection points between distinct bow shocks. Transient bright knots
in the HH~47 bow shock may also arise from Mach stems. The timescales for the appearance
of these knots are consistent with those predicted by the numerical models.
Hence, Mach stems are likely to occur in a variety of astrophysical
situations that involve intersecting shocks, though the Mach stem sizes will not be larger
than the diameter of the obstacles that produce the bows, and stems only form readily
when the obstacles are smaller than the cooling distance behind the shocks. As
a result, most astrophysical Mach stems will be difficult to resolve spatially.

More complex numerical simulations should give further insights into Mach stem formation
and decay.  Relaxing the 2-D symmetry of the simulations will greatly increase the geometrical possibilities,
and will likely reduce the area of Mach stems in most systems.  An interesting case to model
would be when one bow shock overtakes another, as Mach stems should form
over a range of times centered near where the faster bow overtakes the slower one.
Another issue is to consider the role magnetic fields may play in the phenomenon.
Magnetic fields oriented perpendicular to the flow will become bent as 
the flow drags them into the intersection points between shocks, and the resulting tension
force will provide some back pressure that should enhance Mach stems as they lower
the Alfv\'enic Mach numbers.  Magnetic 
fields also lengthen cooling zones, which should raise the effective $\gamma$ of
the system and make it easier for Mach stems to form over larger size scales.
Whether or not the geometry and cooling of a given object
permit Mach stems to form, intersection points
of shocks are natural locations for density enhancements and time-variability in clumpy
supersonic flows.

\acknowledgements
This research is supported by the Department of Energy National Nuclear Security
Administration under Award Number DE-NA0001944 (for operations of the Omega Facility)
and National Laser Users Facility (NLUF) grants DE-NA0002037 and DE-NA0002722 to the PI. 
We would like to thank the staff at LLE for their efficiency and helpfulness with the experiment,
and General Atomics for their expertise with manufacturing the targets used in the experiments.
A helpful referee motivated us to explore Mach numbers more thoroughly and improved
the overall presentation of the paper.

\begin{center}
\begin{table}
\caption{Model, Experimental, and Observational Parameters}
\begin{tabular*}{4.4in}{ccc}
\noalign{\medskip}
\hbox to 1.3in{\hfil Type\hfil}&\hbox to 1.3in{\hfil Mach Number\hfil}&\hbox to 1.3in{\hfil Gamma\hfil}\\
\noalign{\hrule}
\noalign{\smallskip}
Simulations&1.6; 5.2; 30&1.2; 1.4; 1.67\\
Experiments&20&1.42\\
HH Objects&4$-$30\tablenotemark{a,b}&1$-$1.67\tablenotemark{c}\\
\end{tabular*}
\tablenotetext{a}{\ Lower range for jet knots, higher range for bow shocks}
\tablenotetext{b}{\ Alfv\'enic Mach number is lower}
\tablenotetext{c}{\ No fixed $\gamma$ value. $\gamma_{eff}$ = 5/3 for d $\lesssim$ d$_{C}$;
                  $\gamma_{eff}$ = 1 for d $\gtrsim$ d$_C$; d$_C$ $\sim$ 100~AU is the cooling
                  distance}
\end{table}
\end{center}

\begin{center}
\begin{table}
\caption{Critical Angle $\alpha_C$ (degrees) for Mach Stem Formation}
\begin{tabular*}{5.2in}{ccccccc}
\noalign{\medskip}
&\multispan3{\hfil AstroBEAR Simulations\tablenotemark{a}\hfil}&&
 \multispan2{\hfil Analytical Formulae\tablenotemark{b}\hfil}\\
\noalign{\hrule}
\noalign{\smallskip}
\hbox to 1in{\hfil Gamma\hfil}&
             \hbox to 0.5in{\hfil M=1.6\hfil}&\hbox to 0.5in{\hfil M=5.2\hfil}& \hbox to 0.5in{\hfil M=30\hfil}&&
             \hbox to 1in{\hfil dR92\hfil}&\hbox to 0.5in{\hfil CF48\hfil}\\
1.2 &47.1&45.5&44.1&\quad&46.5&56.4\\
1.4 &46.4&43.0&42.9&&40.0&45.6\\
1.67&44.8&33.8&37.3&&35.3&36.8\\
\end{tabular*}
\tablenotetext{a}{\ Typical measurement uncertainties $\pm$ 1$^\circ$}
\tablenotetext{b}{\ dR92: Equation 6 of DeRosa et al. (1992); CF48: Courant and Friedrichs (1948)}
\end{table}
\end{center}

\clearpage

\clearpage

\null

\end{document}